\documentclass[]{article}  
\RequirePackage{amsthm,amsmath,amsfonts,amssymb}
\RequirePackage{graphicx}
\RequirePackage[authoryear]{natbib}
\usepackage{authblk}
\usepackage{epsfig}
\usepackage{graphicx, psfrag}
\usepackage{amsmath}
\usepackage{xfrac} 
\usepackage{amsmath, nccmath}
\usepackage{amssymb}
\usepackage{verbatim}
\usepackage{tikz}
\usetikzlibrary{arrows,calc,shapes,decorations.pathreplacing}
\usepackage{lineno}
\usepackage{xcolor}
\usepackage{natbib}
\usepackage{pdfpages}
\usepackage{xr}
 \usepackage{placeins}
\usepackage[figuresright]{rotating}
\usepackage{lipsum}            

\usepackage[margin=1.25in,footskip=0.25in]{geometry}

\begin{document}

\title{CQUESST: A dynamical stochastic framework for predicting soil-carbon sequestration}

\author[A]{Dan Pagendam}
\author[B]{Jeff Baldock}
\author[C]{David Clifford}
\author[B]{Ryan Farquharson}
\author[D]{Lawrence Murray}
\author[E]{Mike Beare}
\author[E]{Denis Curtin}
\author[F]{Noel Cressie}
\affil[A]{CSIRO Data61, Australia}
\affil[B]{CSIRO Agriculture and Food, Australia}
\affil[C]{Formerly CSIRO}
\affil[D]{Independent Researcher}
\affil[E]{The New Zealand Institute for Plant and Food Research Limited, New Zealand}
\affil[F]{National Institute for Applied Statistics Research Australia, University of Wollongong, Australia}
\maketitle

\begin{abstract}
A statistical framework we call CQUESST (Carbon Quantification and Uncertainty from Evolutionary Soil STochastics), which models carbon sequestration and cycling in soils, is applied to a long-running agricultural experiment that controls for crop type, tillage, and season.    The experiment, known as the Millenium Tillage Trial (MTT), ran on 42 field-plots for ten years from 2000-2010; here CQUESST is used to model soil carbon dynamically in six pools, in each of the 42 agricultural plots, and on a monthly time step for a decade.  We show how CQUESST can be used to estimate soil-carbon cycling rates under different treatments.  Our methods provide much-needed statistical tools for quantitatively inferring the effectiveness of different experimental treatments on soil-carbon sequestration.  The decade-long data are of multiple observation types, and these interacting time series are ingested into a fully Bayesian model that has a dynamic stochastic model of multiple pools of soil carbon at its core.  CQUESST's stochastic model is motivated by the deterministic RothC soil-carbon model based on nonlinear difference equations.  We demonstrate how CQUESST can estimate soil-carbon fluxes for different experimental treatments while acknowledging uncertainties in soil-carbon dynamics, in physical parameters, and in observations. CQUESST is implemented efficiently in the probabilistic programming language {\tt Stan} using its {\tt MapReduce} parallelization, and it scales well for large numbers of field-plots, using software libraries that allow for computation to be shared over multiple nodes of high-performance computing clusters.
\end{abstract}

\section{Introduction.}\label{intro}

Soil contains the largest store of organic carbon in the terrestrial environment, accounting for more than twice that found in vegetation \citep{Scharlemann2014}.  Agricultural soils comprise 37\% of Earth's surface \citep{Smith2008} and have been seen as potential sinks for sequestering atmospheric carbon by altering agricultural management practices.   Practices that conserve or increase the mass of carbon stored in soil are important mitigation strategies to slow down climate change whilst also enhancing agricultural productivity by improving soil fertility, resilience, and sustainability \citep{Baldock1999, Lal2002, Smith2008, Lal2011, Baldock2012, Sommer2014, Minasny2017, Georgiou2022}.

Process-based deterministic models for describing soil organic carbon (SOC) dynamics are ubiquitous in soil-science research.  To predict the potential outcomes of agricultural management practices on SOC stocks, studies have employed deterministic models such as RothC \citep{Jenkinson1990, Li2016}, CENTURY \citep{Parton1993, Nicoloso2020}, DAYCENT \citep{DelGrosso2001, Lemma2021}, APSIM \citep{Luo2014, Oleary2016, Mohanty2020}, EPIC \citep{Causarano2008, Le2018}, DNDC \citep{Li1994, Li2016}, SOMic \citep{Woolf2019}, and ICBM \citep{Andren1997, Bolinder2012}, noting that there are others that have been developed for specific applications \citep[e.g., in forestry; ][]{Black2014, Mao2019}.  It is widely acknowledged that predictions from SOC models are most useful when accompanied by quantification of uncertainty \citep{Ogle2003, Refsgaard2007, Post2008, Juston2010, Clifford2014}.  

To tackle uncertainty in models and estimates, practitioners have on occasions adopted approaches that are statistically questionable or that did not properly quantify the sources of uncertainty.  For example, \cite{Wang2005} and \cite{Juston2010} employed the non-statistical Generalized Likelihood Uncertainty Estimation (GLUE) framework of \cite{Beven1992} and \cite{Beven2001}, an approach that has been criticized for producing ``incoherent and inconsistent'' results \citep{Christensen2004, Mantovan2006, Stedinger2008}.  Other studies, like that of \cite{Andren1997}, \cite{Post2008}, and \cite{Luo2014} employed sensitivity analyses in which model parameters were sampled or perturbed in order to study the resulting variability in model output. Sensitivity analyses can identify important parameters \citep{OHagan2012}, but they are not easily adapted to quantifying uncertainties in predictions of latent (unobservable) processes.  Existing models of soil-carbon cycling in multiple carbon pools do not allow for a satisfactory, statistical accounting for uncertainties in data, parameters, and process dynamics that are needed to confidently make claims about differences between management practices.  

In what is to follow, we present a statistical framework that shows how existing soil-carbon modeling of field trials can be enhanced to account for these uncertainties when assessing the potential of different practices, such as tillage type and use of cover crops, to improve soil-carbon sequestration.  This statistical framework is based on a Bayesian hierarchical statistical model (BHM) that allows prediction of the carbon fluxes (with uncertainties) cycling between various latent soil-carbon pools.  Further, it allows assessment of hypotheses about the pools, their parameters, and how these vary as a function of agricultural treatments, all in a science-driven dynamical setting.  We call our new framework {\em CQUESST} ({\em Carbon Quantification and Uncertainty from Evolutionary Soil STochastics}), which embeds a stochastic, dynamical version of the popular six-pool RothC model \citep{Jenkinson1990, Coleman1996} for soil-carbon dynamics, into a BHM.  

Bayesian hierarchical modeling \citep[BHM; e.g.,][]{Berliner1996, Wikle2007} is a rigorous statistical framework that has gained widespread traction for modeling complex spatio-temporal phenomena in a variety of fields, including climate science \citep[e.g.,][]{Kang2012, Katzfuss2017, Zhang2020, Zammit2022}, oceanography  \citep[e.g.,][]{Wikle2013, Britten2021} and hydrology \citep[e.g.,][]{Pagendam2014, Li2020}, but its potential in soil-carbon modeling has not been fully realized.  At the core of a BHM is the partitioning of a complex, joint-probability distribution into a product of conditional-probability distributions that describe uncertainties in the observed data (via a \emph{data model} at the first level), uncertainties in the underlying scientific process (via a \emph{process model} at the second level), and uncertainties in the parameters (via a \emph{parameter model} or \emph{prior distribution} at the last level).

In our case, the data model quantifies the measurement errors and the combination of the soil-carbon pools that exist in field data; the process model is a science-driven, dynamical statistical process model that describes the evolution of SOC in multiple latent soil-carbon pools through time; and the parameter model captures in a probabilistic manner, beliefs about the values and variability of the parameters in the data model and the process model \emph{prior} to observing the data.

In this article, we demonstrate how the CQUESST framework has a BHM at its core and produces posterior distributions that provide statistical insights into how soil-carbon cycling can change under different agricultural treatments.  Specifically, our study makes use of data from the Millennium Tillage Trial (MTT), a long-running agricultural field trial, that includes a range of tillage and cropping treatments that alter the balance between inputs and losses of organic carbon from soil.  A key aim in our analysis of the MTT with CQUESST, is to assess how decay rates (with uncertainties) of SOC stocks vary as a consequence of the tillage-cropping treatments employed.  This is an important consideration for improving soil-carbon modeling employed in national carbon accounting, where it is often assumed that soil-carbon decay rates are uniform over agricultural management practices.  A secondary aim of our study is to quantify the carbon flux (with uncertainties) from the soil to the atmosphere for each treatment over the decade that the field-trial ran, providing insights into which management practices may have the greatest potential for mitigating climate change over long time-horizons.  Finally, we also show how CQUESST can be used to make inferences on the latent carbon pools of the soil, providing estimates and uncertainty quantification for these over the duration of the MTT.

In recent years, BHMs have been introduced to the soil-carbon-modeling community \citep[e.g.,][]{Cable2009, Clifford2014, Kim2014, Ogle2014, Li2015, Ogle2015, Davoudabadi2021, Davoudabadi2023}.  Our earlier paper \citep{Clifford2014} demonstrates its application on a simple, single-pool model of soil-carbon dynamics.  In what is to follow, this prototype is taken in new directions, particularly the embedding of a multi-pool model that is used in inferring statistically the effect of tillage-cropping treatments on soil-carbon cycling using the Millennium Tillage Trial data.  

In Section \ref{statisticalModel}, we describe the conditional-probability levels of the BHM at the core of CQUESST, used for studying the MTT.  Section \ref{bayesianComputation} gives a brief description of the Bayesian computational methods used in CQUESST, which are then applied to the MTT data in Section  \ref{InferenceResults}.  Our statistical analysis compares carbon sequestration across the various treatments used in the MTT, as well as the soil-carbon decay rates as a function of factors included in the designed experiment.  Section \ref{discussion} discusses the results and the importance of the CQUESST framework for addressing the grand challenge of slowing climate change induced by carbon-based greenhouse gases.  Online Supplemental Material justifies a number of results given in the main text.

\section{CQUESST: A Biophysical-Statistical Model of Soil-Carbon Cycling.}\label{statisticalModel}

In this section, we introduce the Millennium Tillage Trial (MTT) dataset and develop the levels of the BHM used in Section \ref{InferenceResults} to analyze the dataset.

\subsection{The Millennium Tillage Trial Dataset.}\label{MTT}

\begin{figure} 
\begin{center} 
\includegraphics[width=1.0\textwidth]{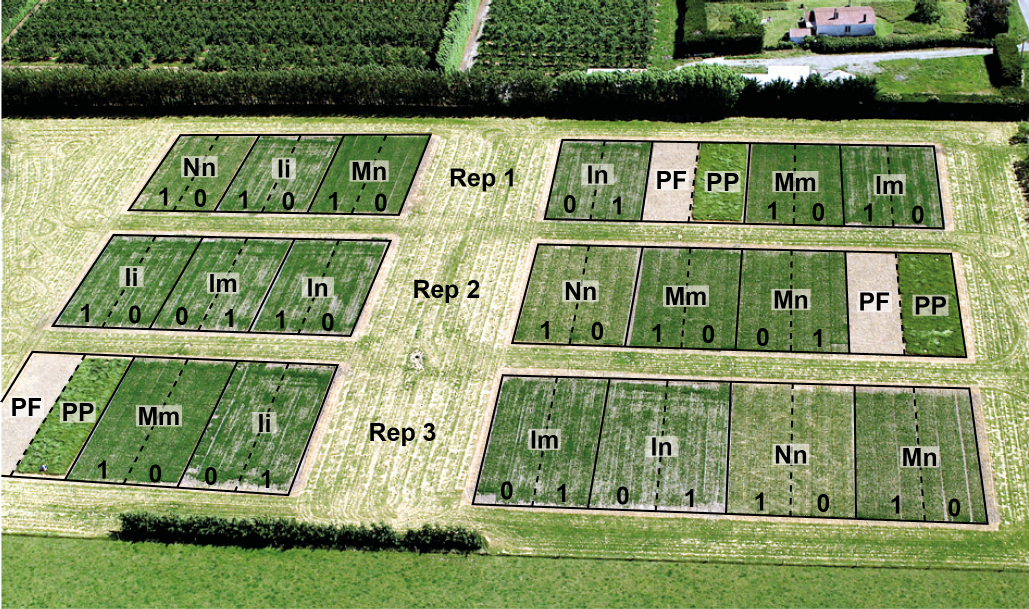}
\caption{Aerial photograph of the Millenium Tillage Trial site in Lincoln New Zealand.  Spring tillage is coded using the capital letters ``N'' (no spring tillage), ``M'' (minimal spring tillage), and ``I'' (intensive spring tillage).  Autumn tillage was coded using the lower-case letters ``n'' (no autumn tillage), ``m'' (minimal autumn tillage) and ``i'' (intensive autumn tillage), respectively.  Tillage treatments were replicated three times, as were two type of no-tillage, namely permanent pasture (PP) and permanent fallow (PF).  The position of the treatments within the replicate were allocated randomly.   \label{MTTAerial}}
\end{center} 
\end{figure}

Long-term agricultural field trials are important sources of experimental data for studying the responses of soil-carbon stocks to different agricultural management practices.  The Millennium Tillage Trial (MTT) was a long-running agricultural trial that took place at Lincoln, New Zealand from 2000 -- 2010.  It was a highly strategic, decade-long field experiment, designed to identify tillage and crop-cover practices for maintaining soil organic carbon (SOC) following the conversion of long-term-pasture to arable cropping. Specifically, the trial examined the effects on soil-carbon sequestration, of spring and autumn tillage treatments and the presence or absence of winter cover crops.  These different tillage-cropping treatments were applied across 42 field-plots, each with dimensions of 9m $\times$ 28m.  The MTT used three levels of a \emph{spring} tillage treatment, listed in increasing order of intensity: no spring tillage (denoted ``N''), minimal spring tillage (``M''), and intensive spring tillage (``I'').  The same three levels of tillage were also applied in \emph{autumn} and denoted by the lower-case letters ``n'', ``m'', and ``i'', respectively.  In addition to the spring and autumn tillages, a third treatment was used in the MTT, namely whether or not a cover crop was grown over winter.  The presence or absence of the winter cover crop, was coded as a binary variable, taking the values ``1'' and ``0'', respectively.  Figure \ref{MTTAerial} shows the layout of the MTT, where each experimental treatment is described by a three-character code: for example, the two field-plots in the bottom right corner would be ``Mm1'' and ``Mm0''.  The MTT did not employ a full factorial design; instead it was constrained so that the autumn tillage was at the same or lower intensity as that applied in the spring (e.g., there are no codes with ``M'' and ``i'').  This reduced the number of possible treatments from 18 down to 12.  Each of the three-character treatment codes was replicated at three field-plots, resulting in a total of 36 field-plots with these types of treatments.  In addition, the MTT also included six no-tillage field-plots: these comprised three plots of a permanent pasture (PP) treatment, and three plots of a permanent fallow (PF) treatment that applied herbicide to keep the plots plant-free.  In all, the MTT had a total of 14 tillage-cropping treatments applied to 42 field-plots.

At each harvest in the trial, measurements were collected of three types of soil-carbon along with the above-ground plant biomass.  The three types were total organic carbon (TOC), particulate organic carbon (POC), and resistant organic carbon (ROC).  Further details of the trial design and conduct can be found in \cite{Baldock2018}.

\subsection{RothC v26.3.}

In this subsection, we show how biophysical knowledge of soil-carbon cycling through multiple carbon pools can be used to construct dynamical models. Many are deterministic in that identical input gives identical output.  However, the uncertainty in processes that govern the cycling should be accounted for, and this we do through statistical models that build on extant biophysical knowledge.  Popular deterministic models for modeling soil carbon in agricultural systems include RothC \citep{Jenkinson1990}, CENTURY \citep{Parton1993}, DAYCENT \citep{DelGrosso2001}, and APSIM \citep{Luo2014}.  Of these models, RothC focuses solely on carbon, whereas CENTURY, DAYCENT, and APSIM model carbon and nitrogen jointly.  The RothC model forms the basis for the soil-carbon component of FullCAM \citep{Richards2004} that is used by the Australian government to produce national greenhouse gas accounts.

RothC is a deterministic, multi-pool model of soil-carbon dynamics; in what follows we describe version, v.26.3 \citep{Coleman1996}.  Within the RothC model, the total mass of carbon in the soil is partitioned into six  scientifically motivated `pools' of carbon, representing substrate material that differ in chemical composition and decomposability (four pools) and biological material in microbial pools (two pools).  

Recall the MTT introduced in Section \ref{MTT}: we now introduce some notation for the analysis that follows in this article.  For month $t$ within field-plot $i$, any addition of carbon to the soil from plant matter is modeled through the time-varying forcing variable, $P_{i,t}$, and the masses of carbon within the six pools are represented as a multivariate stochastic process through the time-varying process vector,

\begin{equation}
{\bf Y}_{i,t} = (D_{i,t}, R_{i,t}, F_{i,t}, S_{i,t}, H_{i,t}, I_{i,t})^{\top},
\end{equation} 

\noindent where $D_{i,t}$ is the carbon stock in decomposable plant material, $R_{i,t}$ is the carbon stock in resistant plant material, $F_{i,t}$ is the carbon stock in fast-decomposing biomass, $S_{i,t}$ is the carbon stock in slow-decomposing biomass, $H_{i,t}$ is the carbon stock in humified organic matter, and $I_{i,t}$ is the carbon stock in inert organic matter that does not decay with time.  The two pools, $S_{i,t}$ and $F_{i,t}$, are distinct from the other pools because these correspond to organic carbon contained in biological microbial material. Finally, we note that field-plot $i$ will have a unique tillage-cropping treatment assigned to it (Figure \ref{MTTAerial}).

The six carbon pools that appear as elements in ${\bf Y}_{i,t}$ cycle amongst each other and the atmosphere and are referred to as `conceptual carbon pools' \citep{Skjemstad2004} or `modeled carbon pools' \citep{Poeplau2013} by soil scientists.  In what is to follow, we call them respectively, the D pool, R pool, F pool, S pool, H pool, and I pool.  Soil chemists cannot directly measure the carbon content of these latent carbon pools.  Instead, observations are made of ``measurable fractions'', subcomponents of the soil that can be separated mechanically (i.e., using sieves) and chemically \citep[e.g.,][]{Baldock2013}.  In the MTT, total organic carbon (TOC), particulate organic carbon (POC), and resistant organic carbon (ROC) were measured.  POC was measured as the organic carbon associated with soil particles larger than 50 $\mu$m.  ROC was measured as the organic carbon associated with polyaromatic structures consistent with, but not necessarily limited to, carbon contained in charcoal using $^{13}$C nuclear magnetic resonance.  In terms of its six pools, RothC recognises them as combinations of the pools:

\begin{align}\label{measureableFractions}
TOC_{i, t} &= D_{i,t} + R_{i,t} + F_{i,t} + S_{i,t} + H_{i,t} + I_{i,t} \notag  \\
POC_{i, t} &=  D_{i,t} + R_{i,t} + F_{i,t} \\
ROC_{i, t} &= I_{i,t} \notag.
\end{align}

Within RothC, there is no distinction made between the measurement and the process it is measuring.  That distinction is important, and it is modeled carefully in the CQUESST framework. Also, although measurements of POC have on occasion been equated solely with $R_{i,t}$ \citep[e.g.,][]{Skjemstad2004}, the actual measurable fraction that is classified as POC also includes decomposable plant matter, $D_{i,t}$, and the fast-decomposing biomass, $F_{i,t}$, which is attached to the plant-material substrate.  The rationale for excluding $F_{i,t}$ and $D_{i,t}$ from $POC_{i, t}$ has been that these pools typically only represent small proportions of the POC.  However, immediately after the addition of plant matter to the soil, $D_{i,t}$ can become elevated.  Furthermore, despite the small contribution of $D_{i,t}$ and $F_{i,t}$ to POC, one should maintain mass-balance and link soil-carbon measurements to their respective carbon pools, as is achieved in Equation (\ref{measureableFractions}).


At each time step, some of the carbon within each of the pools (with the exception of $I_{i,t}$, which is inert and hence remains unchanged over time) undergoes microbial decay and is either: (i) transformed into carbon belonging to one of the pools (either the same or a different type); or (ii) released into the atmosphere as carbon dioxide (CO$_2$) by microbial respiration.  It has been estimated that microbial respiration within soils contributes the largest flux of CO$_2$ from terrestrial ecosystems to the atmosphere \citep{Ogle2015}.  The deterministic dynamics of RothC are presented in Supplemental Material 1; in the following section, we introduce a stochastic analogue of these dynamics that is used in CQUESST.

\subsection{A Dynamical Stochastic Model Based on RothC.}
Here we eschew the deterministic, error-free RothC biophysical representation of soil-carbon dynamics.  Instead, we developed a dynamical stochastic model \emph{based on RothC}, in which we recognize scientific uncertainty.  Stochasticity is introduced into each of the transition equations outlined in Supplemental Material 1, in the form of additive Gaussian noise on the logarithmic scale  \citep{Clifford2014}. 
The stochastic equations, for the multi-pool model considered here have multiplicative errors, $\exp \{ \eta_{X,i,t} \}$, on the natural scale and are:

{\footnotesize
\begin{align}\label{transitionEqStoch}
D_{i, t + \Delta t} &= \exp\{ \log(D_{i,t} e^{-r_{i, t + \Delta t} \frac{K_{D}}{12} \Delta t} + p_{P \rightarrow D} P_{i,t} + p_{M \rightarrow D} M_{i,t} ) + \eta_{D,i,t}\} \notag \\
R_{i, t + \Delta t} &= \exp\{\log(R_{i,t} e^{-r_{i, t + \Delta t} \frac{K_{R}}{12} \Delta t} \notag + (1 - p_{P \rightarrow D}) P_{i,t}  + p_{M \rightarrow R} M_{i,t}) + \eta_{R,i,t}\} \notag \\
F_{i, t + \Delta t} &= \exp\{\log(F_{i,t} e^{-r_{i, t + \Delta t} \frac{K_{F}}{12} \Delta t} + p_{U \rightarrow F} U_{i,t} + p_{V  \rightarrow F}V_{i,t} + p_{M \rightarrow F} M_{i,t}) + \eta_{F,i,t}\} \\
S_{i, t + \Delta t} &= \exp\{\log(S_{i,t} e^{-r_{i, t + \Delta t} \frac{K_{S}}{12} \Delta t} + p_{U  \rightarrow S} U_{i,t} + p_{V  \rightarrow S} V_{i,t})  + p_{M \rightarrow S} M_{i,t}  + \eta_{S,i,t}\} \notag \\
H_{i, t + \Delta t} &= \exp\{\log(H_{i,t} e^{-r_{i, t + \Delta t} \frac{K_{H}}{12} \Delta t} + p_{U  \rightarrow H} U_{i,t}  + p_{V  \rightarrow H} V_{i,t}  + p_{M \rightarrow H} M_{i,t}) + \eta_{H,i,t}\} \notag \\
I_{i, t + \Delta t} &= I_{i,t}, \notag
\end{align}
}

\noindent where
\begin{align}
U_{i,t} &= \sum_{X \in \{D, R, F, S\}} \hspace{-5mm} X_{i,t}(1 - e^{-r_{i, t + \Delta t} \frac{K_{X}}{12}\Delta t}), \label{processNodesStoch} \\
V_{i,t} &= H_{i,t} (1 - e^{-r_{i, t + \Delta t} \frac{K_{H}}{12}\Delta t}). \notag
\end{align}

\noindent In (\ref{transitionEqStoch}) and (\ref{processNodesStoch}), $\Delta t = 1$ month; $K_{D}$, $K_{R}$, $K_{F}$, $K_{S}$, and $K_{H}$ are respectively the annual decay-rates for carbon in the $D$, $R$, $F$, $S$, and $H$ pools for field-plot $i$; $r_{i, t}$ is a decay-rate modifier that is applied to the decay-rates as a result of changes in climate and ground cover; by definition, the I pool does not vary over time; $U_{i,t}$ is the total carbon decayed from $D_{i,t}$, $R_{i,t}$, $F_{i,t}$, and $S_{i,t}$ pools in field-plot $i$, that remained in the soil from month $t$ to $t + 1$; $V_{i,t}$ is the carbon decayed from $H_{i,t}$ in field-plot $i$, that remained in the soil from month $t$ to $t + 1$; and $P_{i,t}$ is the addition of carbon from plant material to the soil in field-plot $i$ and month $t$. The role of each of the dynamic-process model parameters (e.g., $p_{P \rightarrow D}$, $p_{M \rightarrow D}$, etc.) is derived from soil-science considerations and expanded in Supplemental Material 2.  Equation (\ref{transitionEqStoch}) includes the random variables $\eta_{D,i,t} \sim N(-\tfrac{\sigma^2_D}{2}, \sigma^2_D)$; $\eta_{R,i,t} \sim N(-\tfrac{\sigma^2_R}{2}, \sigma^2_R)$; $\eta_{F,i,t} \sim N(-\tfrac{\sigma^2_F}{2}, \sigma^2_F)$; $\eta_{S,i,t} \sim N(-\tfrac{\sigma^2_S}{2}, \sigma^2_S)$; and $\eta_{H,i,t} \sim N(-\tfrac{\sigma^2_H}{2}, \sigma^2_H)$ as independent, normally (i.e., Gaussian) distributed random errors. Here, $N(\mu, \sigma^2)$ denotes a normal distribution with mean $\mu$ and variance $\sigma^2$, and its exponential is a lognormal distribution with mean, $\exp\{ \mu + \sigma^2/2 \}$.  Hence the exponential of $N(-\sigma^2/2, \sigma^2)$ has a mean of 1, and consequently the means of the state equations in (\ref{transitionEqStoch}) agree with the deterministic equations in Supplemental Material 1.  To diagnose the assumption of independence between field-plots shown in Figure \ref{MTTAerial}, we performed a geostatistical analysis of the MTT data.  Our analysis is given in Supplemental Material 3 and shows no evidence of spatial dependence at the scale of the MTT site.

Equations (\ref{transitionEqStoch}) and (\ref{processNodesStoch}) describe a dynamical system where mass-balance is preserved through an atmospheric pool of carbon, released from the soil to the atmosphere in the form of the greenhouse gas CO$_2$.  In Mg C ha$^{-1}$ y$^{-1}$, the \emph{atmospheric flux} of carbon from the $i$th field-plot over the interval $[0, T]$ can be written as 

\begin{equation}
A_{i} = \frac{12}{T} \sum_{X \in \{ D, R, F, S, H, I\} } (X_{i, 0} - X_{i, T}),
\end{equation}

\noindent where $X_{i, T}$ represents the carbon stock in Mg C ha$^{-1}$ for pool $X$ in the final month of the MTT. Negative values of $A_{i}$ represent soil-carbon sequestration (note that $A_i$ is a carbon flux, not a CO$_2$ flux).  In the MTT, tillage-cropping treatment $\tau$ is applied to three field-plots; suppose they are $i(\tau)_1, i(\tau)_2,$ and $i(\tau)_3$.  Then the flux of atmospheric carbon for treatment $\tau$ is simply computed as the area-weighted average of individual fluxes: 

\begin{equation}\label{eq:Atau}
A^{(\tau)} = \frac{\sum_{j = 1}^3 a_{{i(\tau)}_j} A_{{i(\tau)}_j}}{\sum_{j = 1}^3 a_{{i(\tau)}_j}}, 
\end{equation}

\noindent where $a_i$ is the area of field-plot $i$.  Thus, the best tillage-cropping practice is the one that minimizes $A^{(\tau)}$ over all 14 tillage-cropping treatments $tau$; our carbon-flux results for the MTT are given in Section \ref{soilCarbonDynamics}.

\subsection{Measurement Errors in Observed Soil-Carbon Measurable Fractions.}
Over the course of the MTT, laboratory measurements of three types of organic carbon were made: total organic carbon (TOC), particulate organic carbon (POC), and resistant organic carbon (ROC).  The relationships between these soil-carbon measurable fractions and the latent soil-carbon pools is detailed in (\ref{measureableFractions}).  Each of these measurements is subject to measurement error, which we modeled in the following way:

{\footnotesize
\begin{align}\label{datamodel}
 Z_{\textup{TOC}, i, t} | {\bf Y}_{i,t}, \sigma^2_{\textup{TOC} } & \sim LN(\log[D_{i, t} + R_{i, t}  + F_{i, t}  + S_{i, t}  + H_{i, t}  + I_{i, t} ] - \tfrac{\sigma^2_{\textup{TOC}}}{2},  \sigma^2_{\textup{TOC} }), \notag \\
Z_{\textup{POC}, i, t} | {\bf Y}_{i,t}, \sigma^2_{\textup{POC}} & \sim LN(\log[D_{i, t} + R_{i, t} + F_{i, t}] - \tfrac{\sigma^2_{\textup{POC}}}{2},  \sigma^2_{\textup{POC} }) \\
Z_{\textup{ROC}, i, t} | {\bf Y}_{i,t},  \sigma^2_{\textup{ROC} } & \sim LN(\log[I_{i, t}] - \tfrac{\sigma^2_{\textup{ROC}}}{2},  \sigma^2_{\textup{ROC} }) \notag
\end{align}
}

\noindent where $Z_{\textup{TOC} ,  i, t}$, $Z_{\textup{POC} , i, t}$, and $Z_{\textup{ROC} , i, t}$, are the observations of TOC, POC, and ROC in field-plot $i$ and month $t$; recall ${\bf Y}_{i,t} \equiv (D_{i, t}, R_{i, t}, F_{i, t}, S_{i, t}, H_{i, t}, I_{i, t})^{\top}$; and the notation $LN(\mu, \sigma^2)$ refers to a lognormal distribution with parameters $\mu$ and $\sigma^2$.  Here the (measurement) errors are again multiplicative.

As outlined in detail in Section \ref{subsec:priors} below, we formulated priors on measurement-error variance parameters, $\sigma^2_{\textup{TOC} }$, $\sigma^2_{\textup{POC} }$, and $\sigma^2_{\textup{ROC} }$,  based on laboratory measurements of soil carbon across the field-plots, taken just before the commencement of the MTT (see Table S4.3).

\subsection{Soil-Carbon Models as State-Space Models.}
The process model and data model detailed in equations (\ref{transitionEqStoch}) and (\ref{datamodel}), respectively, specify a general class of models known as  \emph{state-space models} \citep[e.g.,][Ch. 7]{Cressie2011}).   A generic linear latent process is:

\begin{equation*}
\mathbf{Y}_{t + 1} = \mathbf{M}_t  \mathbf{Y}_{t} + \mathbf{g}_t + \boldsymbol{\eta}^{\star}_t,
\end{equation*}

\noindent where $\mathbf{M}_t$ is a matrix whose elements are dictated by parameters that govern the temporal evolution of the system, $\mathbf{g}_t$ is an additive linear term that often corresponds to an external ``forcing'' of the process (e.g., the addition of carbon from plant material to the soil), and $\boldsymbol{\eta}^{\star}_t$ is a random vector (often with zero mean and diagonal covariance matrix) that makes the process stochastic.  For example, the $i$th element $\eta^{\star}_{t, i} \sim N(0, \sigma_{\eta, i}^2)$, independently of the other elements of $\boldsymbol{\eta}^{\star}_t$.  A state-space model also requires the specification of a data model that links observed quantities to the latent-process dynamics.  A generic data model is:

\begin{equation*}
\mathbf{Z}_{t} = \mathbf{H}_t  \mathbf{Y}_{t} + \boldsymbol{\epsilon}^{\star}_t,
\end{equation*}

\noindent where $\mathbf{Z}_{t}$ is a vector of observed quantities, $\mathbf{H}_t$ is a given matrix that links each observed quantity to some linear combination of the state variables in $\mathbf{Y}_{t}$, and $\boldsymbol{\epsilon}^{\star}_t$ is a random vector (often with zero mean and diagonal covariance matrix) that represents measurement error.  For example, the $i$th element $\epsilon^{\star}_{t, i} \sim N(0, \sigma_{\epsilon,i}^2)$, independently of the other elements of $\boldsymbol{\epsilon}^{\star}_t$.

However, our component models of the BHM in CQUESST are not linear.  The framework has at its core a difference equation that leads to a non-linear state-space model, different from the generic linear one just described.  Define the $6 \times 6$ propagator matrix, $\mathbf{M}_{t} \equiv [\mathbf{M}_{1,t}, \mathbf{M}_{2,t}]$, where

\begin{equation*}
\mathbf{M}_{1,t} = \left( { \begin{array}{ccc}
e^{-K_D \Delta_t} & 0 & 0  \\
0 & e^{-K_R \Delta_t} & 0  \\
p_{U \rightarrow F}(1 - e^{-K_D \Delta t}) & p_{U \rightarrow F}(1 - e^{-K_R \Delta t}) & e^{-K_F \Delta_t} + p_{U \rightarrow F}(1 - e^{-K_F \Delta t})\\
p_{U \rightarrow S}(1 - e^{-K_D \Delta t}) & p_{U \rightarrow S}(1 - e^{-K_R \Delta t}) & p_{U \rightarrow S}(1 - e^{-K_F \Delta t})\\
p_{U \rightarrow H}(1 - e^{-K_D \Delta t}) & p_{U \rightarrow H}(1 - e^{-K_R \Delta t}) & p_{U \rightarrow H}(1 - e^{-K_F \Delta t})\\
0 & 0 & 0 \end{array}}\right)
\end{equation*}

\begin{equation*}
\mathbf{M}_{2,t} = \left( { \begin{array}{ccc}
 0 & 0 & 0 \\
 0 & 0 & 0 \\
 p_{U \rightarrow F}(1 - e^{-K_S \Delta t}) & 0 & 0 \\
e^{-K_S \Delta_t} + p_{U \rightarrow S}(1 - e^{-K_S \Delta t}) & p_{U \rightarrow S}(1 - e^{-K_H \Delta t}) & 0 \\
p_{U \rightarrow H}(1 - e^{-K_S \Delta t}) & e^{-K_H \Delta_t} + p_{U \rightarrow H}(1 - e^{-K_H \Delta t}) & 0 \\
0 & 0 & 1\end{array}}\right);
\end{equation*}

\noindent define the six-dimensional vector of carbon inputs to the system as,

\begin{equation*}
\mathbf{g}_{t} \equiv \left( { \begin{array}{c}
p_{P \rightarrow D}P_t + p_{M \rightarrow D}M_t \\
p_{P \rightarrow R}P_t + p_{M \rightarrow R}M_t \\
0 \\
0 \\ 
0 \\
0 \end{array}} \right);
\end{equation*}

\noindent and, corresponding to the data collected, define the observation matrix as,

\begin{equation*}
\mathbf{H}_{t} = \left( { \begin{array}{cccccc}
1 & 1 & 1 & 0 & 0 & 0\\
0 & 0 & 0 & 0 & 0 & 1\\
1 & 1 & 1 & 1 & 1 & 1
\end{array}} \right).
\end{equation*}

The process model and data model together specify a \emph{hierarchical statistical model}, with which we could employ Bayesian or empirical Bayesian statistical methods for inference on parameters and prediction of latent-process dynamics.  From Section \ref{intro}, a BHM, has a third level of the hierarchy given by the prior distribution of parameter vector $\boldsymbol{\theta}$; see Section \ref{subsec:priors}.  Rather than taking an empirical hierarchical modeling (EHM) approach by estimating $\boldsymbol{\theta}$ from the data and ``plugging'' it into the prediction equations, CQUESST uses a BHM, and all inferences come from the posterior distribution of the ``unknowns'' given the data.

\subsection{CQUESST as a State-Space Model.}
On the natural scale of fluxes, the stochastic soil-carbon model specified in (\ref{transitionEqStoch}) can be considered ``almost'' linear, but with multiplicative errors.  That is, the six-dimensional state $\{ \mathbf{Y}_t \}$, evolves dynamically as, 

\begin{equation*}
\mathbf{Y}_{t + 1} = (\mathbf{M}_t  \mathbf{Y}_{t} + \mathbf{g}_t) \odot \boldsymbol{\eta}_t,
\end{equation*}

\noindent where $\boldsymbol{\eta}_t$ is a six-dimensional random vector with $k$th element, $\eta_{t, k} \sim LN(-\frac{\sigma_{\eta, k}^2}{2}, \sigma_{\eta, k}^2)$, and $\odot$ is the Hadamard (elementwise) product of two vectors.  Equivalently, this process model can be written as evolving nonlinearly on the log-scale, with additive Gaussian-process noise:

\begin{equation*}
\log(\mathbf{Y}_{t + 1}) = \log(\mathbf{M}_t  \mathbf{Y}_{t} + \mathbf{g}_t) + \boldsymbol{\eta}^{\star}_t,
\end{equation*}

\noindent where $\boldsymbol{\eta}^{\star}_t$ is a six-dimensional random vector with $k$th element, $\eta_{t, k} \sim N(-\frac{\sigma_{\eta, k}^2}{2}, \sigma_{\eta, k}^2)$, independently of the other elements. Similarly, the data model specified in (\ref{datamodel}) is ``almost'' linear, but with multiplicative measurement error:

\begin{equation*}
\mathbf{Z}_{t} = (\mathbf{H}_t  \mathbf{Y}_{t}) \odot \boldsymbol{\epsilon}_t,
\end{equation*}

\noindent where $\boldsymbol{\epsilon}_t$ is a three-dimensional random vector with $k$th element, $\epsilon_{t, k} \sim LN(-\frac{\sigma_{\epsilon, k}^2}{2}, \sigma_{\epsilon, k}^2)$.  Again, this can be written on the logarithmic-scale as a non-linear model with additive Gaussian measurement error:

\begin{equation*}
\log(\mathbf{Z}_{t}) = \log(\mathbf{H}_t  \mathbf{Y}_{t}) + \boldsymbol{\epsilon}^{\star}_t,
\end{equation*}

\noindent where $\boldsymbol{\epsilon}^{\star}_t$ is a three-dimensional random vector with $k$th element, $\epsilon_{t, k} \sim N(-\frac{\sigma_{\epsilon, k}^2}{2}, \sigma_{\epsilon, k}^2)$, independently of the other elements.

These models are not of a standard form that would yield to maximum likelihood estimation and Kalman Filtering.  Rather than developing approximate solutions that are biased and imprecise, the soil-carbon state-space model in CQUESST given by Equation (\ref{transitionEqStoch}), is embedded into a BHM by including a prior distribution on unknown parameters $\boldsymbol{\theta}$.  Doing so has three substantial benefits: (i) specifying prior distributions allows one to draw upon all sources of information available (including expert opinion and past studies; see Section \ref{priorSection}); (ii) no linear approximations to the process dynamics are needed; and (iii) inference on parameters and state variables can be undertaken simultaneously from the posterior distribution, here through Markov Chain Monte Carlo (MCMC), as in Section \ref{bayesianComputation}.

\subsection{Priors on Model Parameters and Initial Conditions.}\label{priorSection}\label{subsec:priors}

Prior distributions were placed on the initial values in the carbon pools (at month $t = 0$) and on all of the parameters in the model outlined in (\ref{transitionEqStoch}).  This was done in consultation with soil scientists on the project team and drew upon their expert knowledge of how carbon cycles in soil.  For simplicity, we denote the complete vector of parameters as $\boldsymbol{\theta}$, which includes all of the scalar parameters listed in Tables S4.1, S4.2, and S4.3 of Supplemental Material 4.  It is also necessary to specify prior distributions for the initial states of the latent state variables (soil-carbon pools in each of the field-plots), and these are listed in Table S4.4 of Supplemental Material 4.  The prior distributions listed in these tables summarize our beliefs about the likely values for each parameter before observing the MTT data, and they could be broadly classified as either informative or weakly informative.  Furthermore, to assess the sensitivity of our results to the specified priors, we conducted a sensitivity analysis over particularly important parameters, the results of which are reported in Supplemental Material 5.

\subsection{Analyzing a Designed Experiment with CQUESST: The Millennium Tillage Trial.}\label{analysisMTT}

The CQUESST framework can be used to infer model parameters and soil-carbon trajectories for the MTT, where uncertainty in data, process, and parameters are coherently accounted for.  An important scientific question surrounding this dataset is: Do tillage-cropping treatments affect the decay rate of soil carbon?  Soil scientists are aware that crop types and tillage treatments affect the amount of plant material that enters the soil (the main route for soil-carbon sequestration), but they may also affect soil microbial communities responsible for its decomposition.  They may also influence the chemical composition and bioavailability of organic carbon cycling through the D, R, F, and S pools, and thus the rate at which decomposition occurs.  We have particular interest in determining whether the tillage-cropping practices induced differences in the decay rates of the D, R, and H pools, a question that can be studied by allowing the decomposition rates to be functions of the treatments.  We generalize the process model so that the rate parameters in (\ref{transitionEqStoch}), $\{ K_X: X = D, R, F, S, H \}$, depend on the tillage-cropping treatment, $\tau$, through its assignment to field plot $i$, which we have denoted $i(\tau)$.  We write

\begin{equation}\label{eq:alpha}
K_{X, i(\tau)} = \kappa_{X} \alpha_{\tau},
\end{equation}

\noindent where in the process model (\ref{transitionEqStoch}), $i$ is replaced with $i(\tau)$ and $K_X$ is replaced with $K_{X, i(\tau)}$ for carbon $X \in \{ D, R, F, S, H \}$ and treatments, $\tau \in \mathcal{T} = \{ \textup{PP}, \textup{PF}, \textup{Nn0}, \dots, \textup{Ii1} \}$.  In (\ref{eq:alpha}), $\kappa_{X}$ is the marginal decay rate for pool $X$, and $\alpha_{\tau} \in (0, \infty)$ is the treatment-specific modifier applied to decay rates for treatment $\tau$.  Within each treatment, the decay rates, $K_{X, \tau}$, have the desirable property that the scientifically justified ordering of decay rates for the pools is retained.  Diffuse prior distributions on $\{ \alpha_{\tau} : \tau \in \mathcal{T}\}$ are outlined in Supplemental Material 4.

\section{Bayesian Inference and Computation within the CQUESST Framework.}\label{bayesianComputation}

Here we demonstrate how CQUESST can be used to model soil-carbon data from long-term field trials and to improve our understanding of important latent biogeochemical processes that drive carbon cycling.  This is demonstrated in three ways: (i) CQUESST is fitted to the data from the MTT to demonstrate its ability to infer model parameters, latent-process dynamics, and carbon fluxes under different experimental treatments; (ii) uncertainties are captured around these quantities; and (iii) the parameterization of the model can be augmented to see whether carbon-cycling varies under different treatments.  In (iii), we specifically evaluate a hypothesis that soil-carbon decay rates are \emph{not} homogeneous across treatments and will vary depending on the type of production used (e.g., single cropping versus double cropping).

\subsection{Assembling the CQUESST Framework.}

The BHM, which is at the core of CQUESST, uses conditional probability distributions to simplify the complex joint probability distributions encountered in spatio-temporal modeling.  In what follows, we denote the series of observations taken at various times in field-plot $i \in \{1, 2, \dots, 42 \}$ as the vectors $\mathbf{Z}_{\textup{POC}, i}$, $\mathbf{Z}_{\textup{ROC}, i}$, and $\mathbf{Z}_{\textup{TOC}, i}$, which we then concatenate for field-plot $i$ into the vector $\mathbf{Z}_i \equiv (\mathbf{Z}_{\textup{POC}, i}^{\top}, \mathbf{Z}_{\textup{ROC}, i}^{\top}, $  $\mathbf{Z}_{\textup{TOC}, i}^{\top})^{\top}$.  From the MTT data, we wish to obtain estimates of the latent process and parameters and their uncertainties, which in the CQUESST framework comes from the posterior distribution given by an application of Bayes' Rule.  The posterior density function is:

\begin{equation}\label{bayesRule}
p(\mathbf{Y}, \boldsymbol{\theta} | \mathbf{Z}) = \frac{p(\mathbf{Y}, \boldsymbol{\theta}, \mathbf{Z})}{p(\mathbf{Z})} =  \frac{p(\mathbf{Z} | \mathbf{Y}, \boldsymbol{\theta})p(\mathbf{Y} | \boldsymbol{\theta}) p(\boldsymbol{\theta})}{p(\mathbf{Z})},
\end{equation}

\noindent where $\mathbf{Y} \equiv (\mathbf{Y}^{\top}_{0, 1}, \dots , \mathbf{Y}^{\top}_{T, 1}, \dots , \mathbf{Y}^{\top}_{0, 42}, \dots, \mathbf{Y}^{\top}_{T, 42})^{\top}$ is a vector containing all six of the state variables across all 108 months between October 2000 and September 2009, and all 42 field-plots; $\mathbf{Z} \equiv (\mathbf{Z}^{\top}_1,  \dots , \mathbf{Z}^{\top}_{42})^{\top}$ is the vector containing all of the soil-carbon observations across all field-plots and observation times; and $\boldsymbol{\theta}$ is the vector of all model parameters outlined in Tables S4.1, S4.2, and S4.3.  In (\ref{bayesRule}), $p(\mathbf{Y}, \boldsymbol{\theta}, \mathbf{Z})$ is the joint probability density function of the latent soil-carbon processes, the parameters, and the data.  In the BHM, this can be written as the product of three conditional probability density functions, introduced earlier in this paper as the \emph{data model}, the \emph{process model}, and the \emph{parameter model}.  The data model is defined by (\ref{datamodel}), the process model is defined by (\ref{transitionEqStoch}), and the parameter model (or prior) is given by Tables S4.1 - S4.3, with statistical independence between parameters assumed.

The right-hand side of (\ref{bayesRule}) has a normalizing constant $p(\mathbf{Z})$, which is the marginal probability density function of $\mathbf{Z}$ and is generally intractable.  Instead, we use a Markov Chain Monte Carlo (MCMC) algorithm to sample from $p(\mathbf{Y}, \boldsymbol{\theta} | \mathbf{Z})$.  Specifically, we use an extension of Hamiltonian Monte Carlo (HMC), called the No-U-Turn Sampler (NUTS) \citep{GelmanHoffman2014.1}, which is implemented as an auto-tuning, `turnkey' sampling algorithm in {\tt Stan}.  Further, {\tt MapReduce} functionality is available in {\tt Stan} for parallelizing computation of $p(\mathbf{Y}, \boldsymbol{\theta} | \mathbf{Z})$ over numerous CPUs on a high-performance computing cluster.  In the present application, the HMC in CQUESST, parallelizes the posterior calculation of individual field-plots.  In total, we obtained samples from six independent Markov chains, using 20,000 samples for warm-up that were subsequently discarded, and then 50,000 samples for posterior inference in each of these.  For more precise inferences, these were further thinned by taking every tenth sample, resulting in 5,000 approximately independent samples for each of the six chains.  The {\tt Stan} code employing {\tt MapReduce} is available at {\tt https://github.com/dpagendam/CQUESST}.  When performing Bayesian inference with an MCMC algorithm, it is important to verify that the Markov chain provides a representative set of samples from the posterior distribution $p(\mathbf{Y}, \boldsymbol{\theta} | \mathbf{Z})$, and details of the methods used are provided in Supplemental Material 6.

\section{The Millennium Tillage Trial: Inference of Key Components.}\label{InferenceResults}
We now use the CQUESST framework described in Sections \ref{statisticalModel} and \ref{bayesianComputation} to analyze the MTT soil-carbon data and demonstrate how it can be used to make inference on the ``unknowns''.  Specifically, we: (i) fit science-driven dynamics from observed data (despite the fact that each carbon pool is latent and cannot be measured directly); (ii) incorporate prior knowledge around parameters of the model, including those describing the evolution of the soil-carbon trajectories through time; and (iii) estimate and put uncertainty bounds on the quantities in (i) and (ii) using the MCMC samples.  Importantly, the Markov chains for all parameters showed strong evidence of convergence to their stationary distributions, and we refer the reader to Supplementary Material 6 for details.

\subsection{Posterior Inference for Model Parameters.}

Figures S7.1, S7.2, and S7.3 in Supplemental Material 7 show prior and posterior densities for parameters in CQUESST.  Detailed discussion of the Bayesian learning seen with these parameters is given there.  Prior distributions (in gray) and posterior distributions (in purple) that are near to each other indicate that the observed MTT data did not contain information that greatly changed the prior belief about a parameter.  When there is a substantial difference between the (gray) prior and (purple) posterior distributions, or when the prior distribution is so far from the posterior distribution that the prior does not appear in the figure, information in the MTT dataset has drastically changed our prior belief about the plausible values of that parameter.  

In some of the cases, the posterior distributions have shifted away from the prior (e.g., $\kappa_D$ and $\kappa_R$). Parameters where there were less-substantial differences between prior and posterior distributions include the decay-rate parameters $\kappa_F$, $\kappa_S$, and $\kappa_H$, whilst for many of the parameters (see Figure S7.3) that govern the proportions of decomposed carbon that are routed to other pools, there was no apparent learning.  We show that the posterior distribution for $\kappa_D$ favors slightly lower values than the prior distribution, the posterior distribution for $\kappa_R$ favours larger values than the prior distribution; and $\kappa_F$, $\kappa_S$ and $\kappa_H$ show little change.  In the case of $\kappa_F$ and $\kappa_S$ this is unsurprising, since the observations, which relate to the sum of multiple pools, would not be expected to offer much information about the dynamics of the relatively small (in terms of carbon stock) F and S pools.  The similarity between the prior and posterior distributions for $\kappa_H$ is likely due to the fact that: (i) this is typically the largest pool of carbon in the soil by mass, and it's rate of decomposition is well-documented; and (ii) the slow rate of decay of the H pool relative to the duration of the MTT may reduce the information content of the data about this parameter.

From Figure S7.2, it is clear that many of the posterior distributions for variance parameters (process error and measurement errors) shifted away from their priors and that these shifts were more substantial for measurement error variances than for process-error variances.  For the measurement-error variances corresponding to ROC, posteriors remained in roughly the same location, whereas measurement-error variances shifted to lower values for TOC, and to higher values for POC.

\subsection{Soil-Carbon Dynamics.}\label{soilCarbonDynamics}

\begin{figure} 
\begin{center} 
\includegraphics[width=0.85\textwidth]{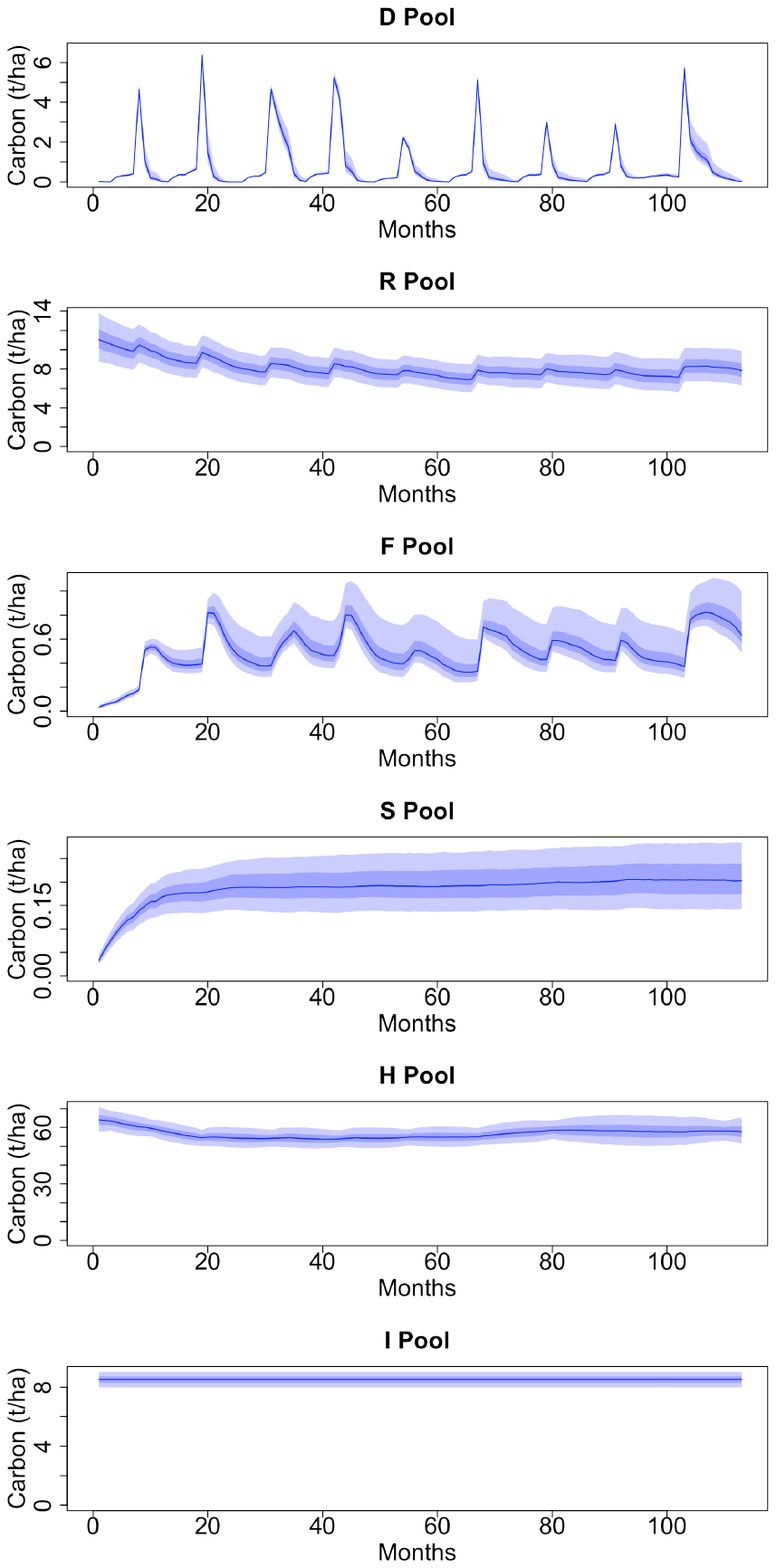}
\caption{Sampled trajectories of latent soil-carbon pools in the second field-plot with tillage treatment Mn0.  Central blue lines show the posterior median, dark blue ribbons span the 50\% posterior two-sided intervals, and light blue ribbons span the 90\% posterior two-sided intervals.  The six soil-carbon pools are: decomposable plant matter (D), resistant plant matter (R), fast (F) and slow (S) decomposers, humus (H), and inert (I) material. \label{fig:Pools_MN0_Field2}}
\end{center} 
\end{figure}

\begin{figure} 
\begin{center} 
\includegraphics[width=0.6\textwidth]{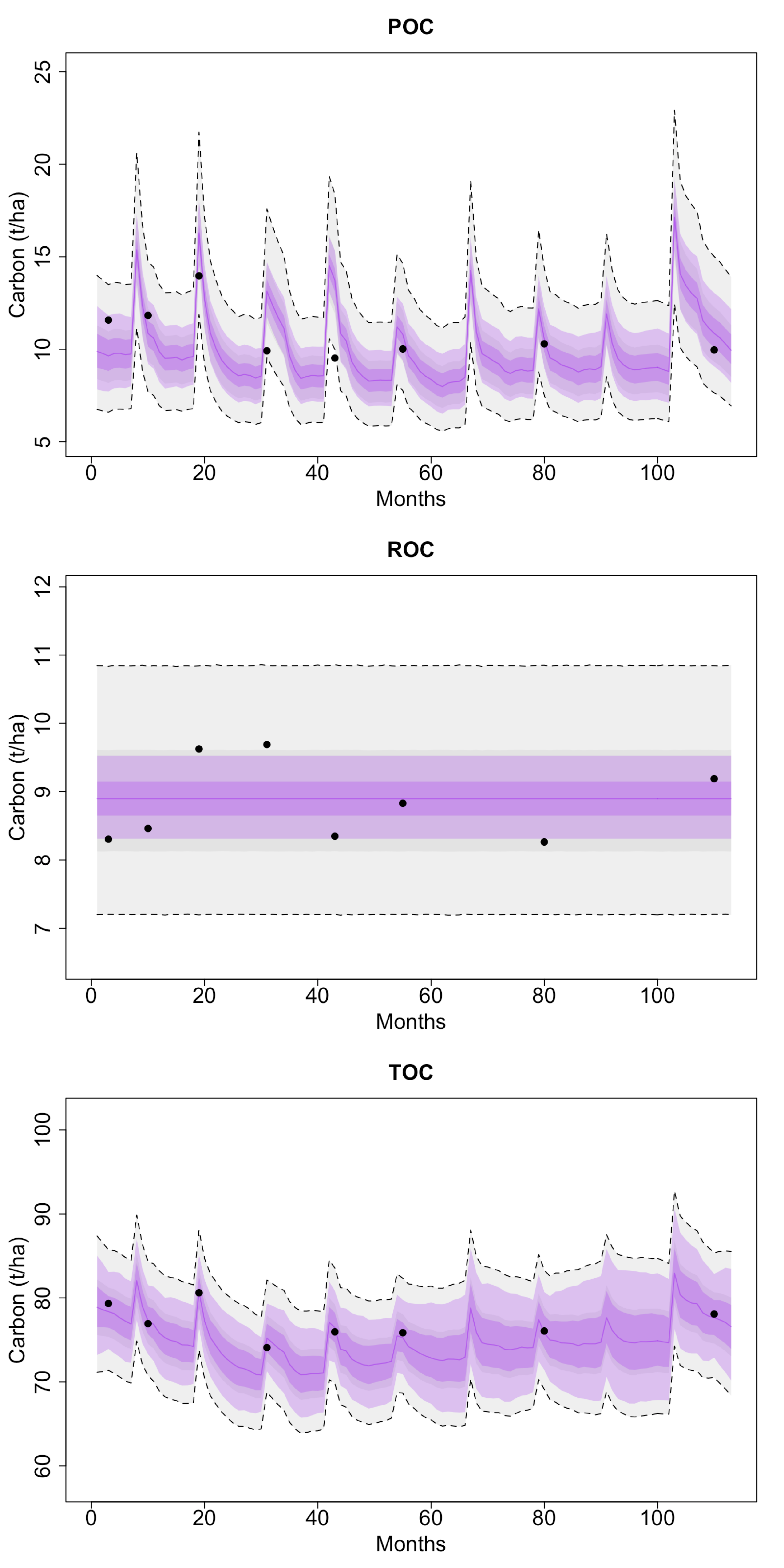}
\caption{Sampled trajectories for the combined, latent soil-carbon pools for each observation type (see equation (\ref{measureableFractions})) in the second field-plot with tillage treatment Mn0.  Black circles show observations, central purple lines show the posterior median of the combined latent pools, dark purple ribbons span the 50\% posterior two-sided intervals of the combined latent pools, and light purple ribbons span the 90\% posterior two-sided intervals of the combined latent pools.  Dashed lines show the 90\% posterior two-sided intervals for the observations with the presence of measurement error taken into account.\label{fig:Observed_MN0_Field2}}
\end{center} 
\end{figure}

Sampled trajectories of soil-carbon pools from the posterior distribution were used to construct Figures \ref{fig:Pools_MN0_Field2} and \ref{fig:Observed_MN0_Field2}.  Figure \ref{fig:Pools_MN0_Field2} shows how CQUESST is able to make inferences on the latent soil-carbon pools given in (\ref{transitionEqStoch}), even though the observations were on aggregated subsets of these given in (\ref{datamodel}).  Plant material ($P_{i,t}$ in (\ref{transitionEqStoch})) enters the D and R pools of a field-plot, and the relative quantities of each in the plant material is determined by the ratio $r_{DPM/RPM}$, where DPM stands for decomposable plant material and RPM stands for resistant plant material.  Further information about $r_{DPM/RPM}$ can be found in Supplemental Materials 2 and 4 and it defines the quantity

\begin{equation*}
p_{P \rightarrow D} = \frac{r_{DPM/RPM}}{1 + r_{DPM/RPM}},
\end{equation*}

\noindent that appears in (\ref{transitionEqStoch}).  In Figure \ref{fig:Pools_MN0_Field2}, we see that the D pool rises and falls each time a crop is sown as plant material enters the soil and is rapidly decomposed.  Similarly, we observe small peaks in the R pool over time that correspond to resistant plant material entering the soil and decomposing more slowly.  The F and S pools start at low masses at the beginning of the MTT and rise over its duration because of the continual input of plant-material substrate for the microorganisms that comprise these pools.  Overall, the F and S pools only represent a small amount of the total carbon stock. We observe that the H pool varies slowly over time, since this pool has a slow decay rate and remains relatively stable.  Finally, the I pool remains completely stable over time, since this pool represents inert carbon in the soil. 

From Figure \ref{fig:Observed_MN0_Field2}, we see that the observed data agree well with the aggregated pools to which the observations correspond.  POC and TOC measurements include material that is in the D and R pools and can therefore exhibit peaks associated with the annual addition of plant material from crops into the soil.  In contrast, the measurements of ROC vary little, since these are solely measurements of the inert I pool that remains stable over time.  Note that the trajectory for TOC is the sum of all trajectories in Figure \ref{fig:Pools_MN0_Field2}.  This can be used to make inferences about how much carbon was accumulated (negative flux) or lost to the atmosphere (positive flux) over the course of the MTT.  Plots of the posterior distribution of the total soil-carbon flux, $A^{(\tau)}$ given by (\ref{eq:Atau}), for each of the 14 tillage-cropping treatments in the MTT are shown in Figure \ref{fig:carbon_change}.  These are discussed in more detail below.

\subsection{Answering Carbon-Sequestration Questions with CQUESST.}

\begin{figure} 
\begin{center} 
\includegraphics[width=1.0\textwidth]{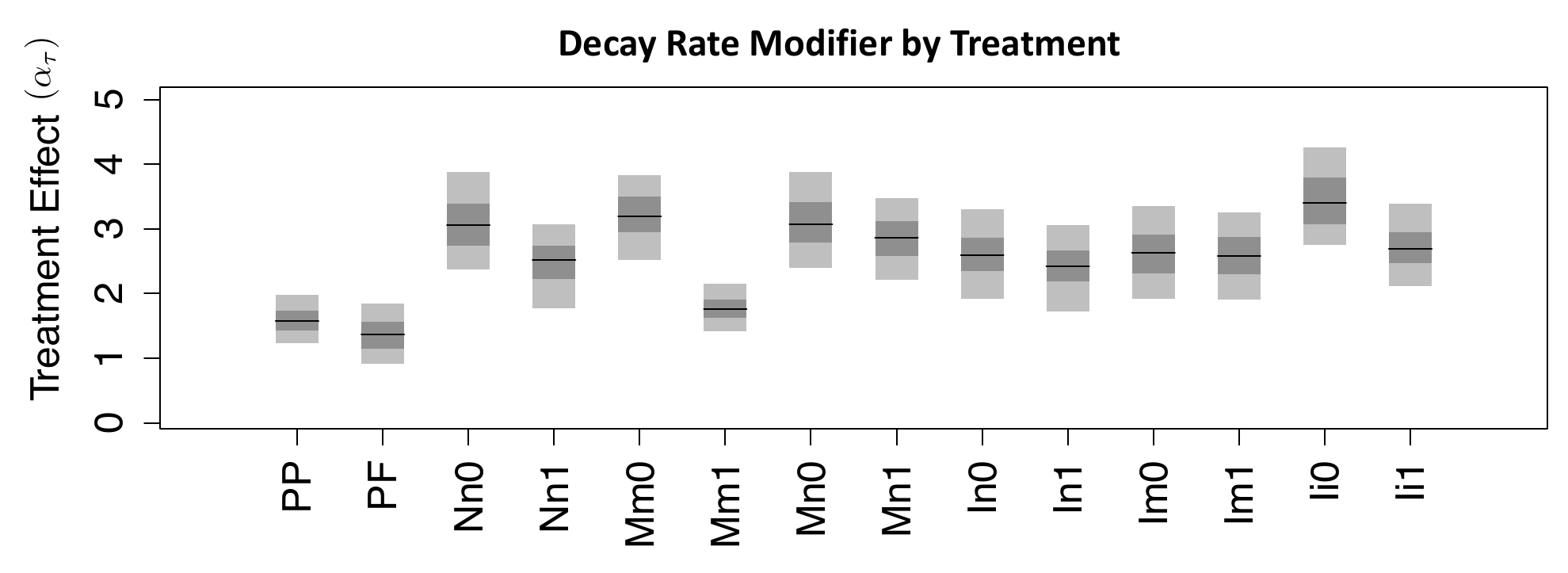}
\caption{Posterior distributions for the multiplicative decay-rate modifier ($\alpha_{\tau}$) for each treatment, $\tau$, in the Millennium Tillage Trial.  The dark horizontal lines show the posterior median, darker rectangles span the 50\% posterior two-sided interval, and lighter coloured rectangles span the 90\% posterior two-sided interval.  \label{fig:alphas}}
\end{center} 
\end{figure}

Posterior distributions for multiplicative treatment effects, $\alpha_{\tau}$, defined in (\ref{eq:alpha}), are shown in Figure \ref{fig:alphas}.  We note that there is substantial variability between the treatments, indicating that soil-carbon decay-rates are not uniform under all tillage-cropping management practices.  This is in contrast to how soil-carbon dynamics are typically modeled in practice and indeed are modeled in the deterministic RothC outlined in Supplemental Material 1.  Of particular note in our results are the no-tillage PF and PP treatments, which are both at the lower range of values for $\alpha_{\tau}$, indicating that decay rates under these two treatments are low.  

\begin{figure} 
\begin{center} 
\includegraphics[width=1.0\textwidth]{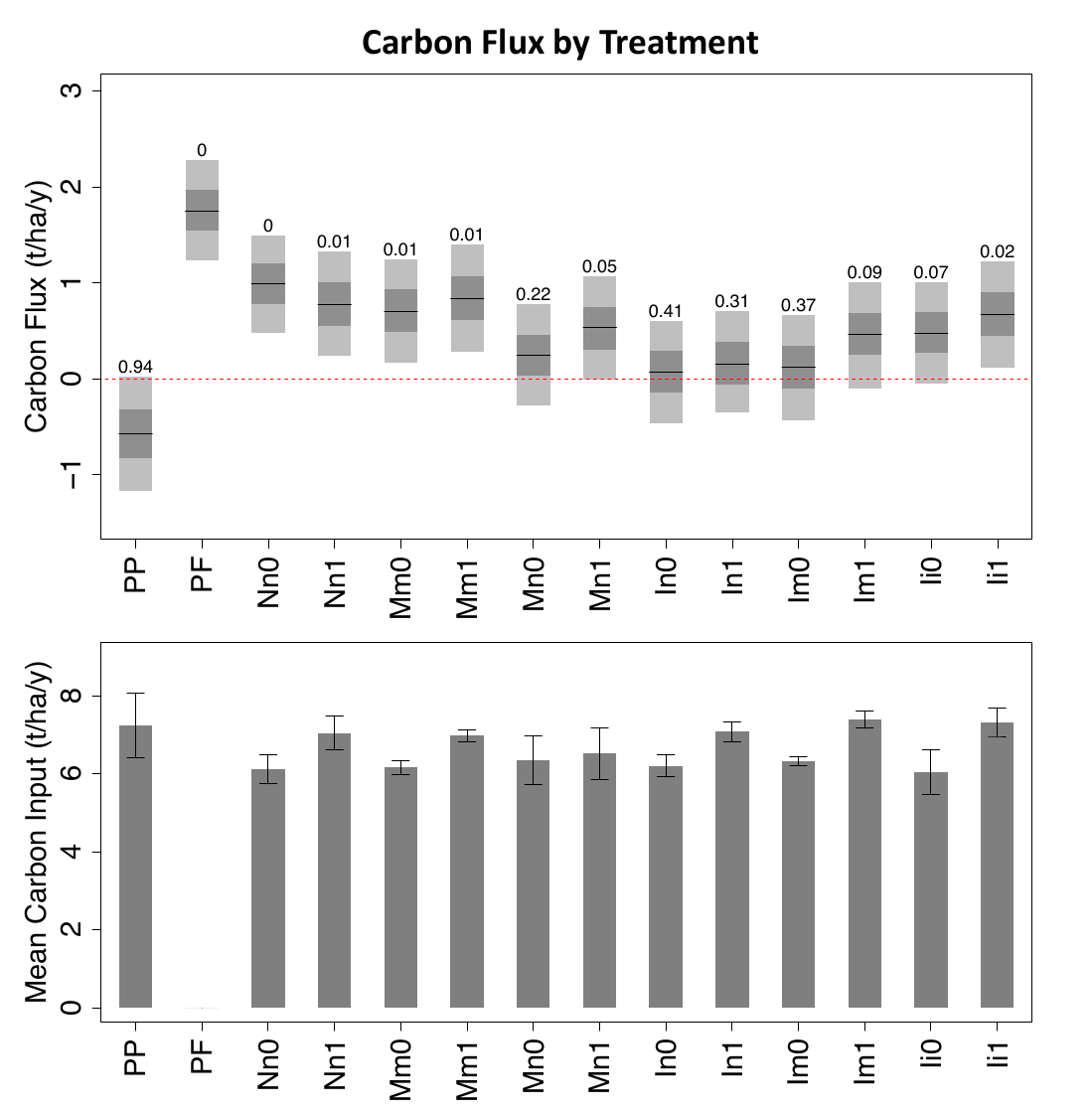}
\caption{Top panel: posterior distribution of fluxes of soil-carbon between the start and end of the Millennium Tillage Trial, averaged over three field-plots for each treatment (see equation (\ref{eq:Atau})). Negative values indicate carbon sequestration.  The dark horizontal lines show the posterior median, darker rectangles span the 50\% posterior two-sided interval, and lighter coloured rectangles span the 90\% posterior two-sided interval.  Numbers above rectangles give the posterior probability that the atmospheric carbon flux for each treatment was less than zero (i.e., soil-carbon sequestration).  Bottom panel: mean annual carbon inputs to the soil from plant material in the MTT (error bars shows $\pm 1$ standard deviation, showing inter-field-plot variability). The three-character codes for treatments are outlined in Figure \ref{MTTAerial}. \label{fig:carbon_change}}
\end{center} 
\end{figure}

Figure \ref{fig:carbon_change} shows that this multiplicative effect does not necessarily dictate the carbon-sequestration potential of a treatment, since sequestration also depends upon the amount of plant material cultivated, with root material (and potentially stubble) entering the soil (see bottom panel of Figure \ref{fig:carbon_change}).  We observe that for the permanent pasture (PP) treatment, there is a high probability of negative carbon flux (i.e., low probability of losing soil carbon to the atmosphere).  In contrast, permanent fallow (PF), which is a chemical fallow with no addition of plant material to the soil, showed very strong evidence of positive carbon flux (i.e., poor carbon-sequestration).  Of the other MTT treatments, the intensive-tillage treatments in the spring, In0, In1, and Im0, showed the next highest probabilities of soil-carbon sequestration; and consistently, no-tillage (Nn0 and Nn1) and minimum-tillage (Mm0 and Mm1) treatments showed low probabilities of sequestration.  Figure \ref{fig:carbon_change} suggests that active tillage treatments that employ a rotation of intensive tillage in the spring, can have greater soil-carbon sequestration potential than those that employ no-tillage or minimum tillage in the spring.

\section{Discussion.}\label{discussion}

The Bayesian hierarchical statistical model that we have implemented through the CQUESST framework provides a powerful tool to better understand a phenomenon of national and ultimately global importance, namely the cycling of carbon stocks in agricultural soils.  CQUESST provides soil scientists with a tool that can be used for combining data, expert knowledge, and biogeochemical-process dynamics in a statistically rigorous way.  In addition, we have shown how CQUESST can also harness all of these sources of information when analyzing designed experiments such as the MTT.  An important attribute of using a BHM is that it allows these inferences to be carried out in the presence of uncertainty in observed data, in our understanding of the process dynamics, and in the model's parameters.  These uncertainties are critical to acknowledge in order to draw statistically rigorous scientific conclusions.  Whilst CQUESST has been developed for the application of soil-carbon modeling, the Bayesian hierarchical modeling approach has many other potential applications in agriculture where mechanistic process models are employed \citep[e.g., crop production models such as APSIM;][]{Holzworth2014}.

Data from tillage trials such as the MTT can be analyzed using simpler statistical approaches such as linear models and analysis of variance (ANOVA). In Supplemental Material 8, we show that results obtained from CQUESST are relatively consistent with those of these simpler statistical approaches.  However, it is important to highlight that CQUESST offers the ability to make more detailed and specific inferences on latent carbon pools based on observations of multiple types of organic carbon.  As examples, the posterior distributions obtained over the latent pools' dynamics (see Figure \ref{fig:Pools_MN0_Field2}) and treatment-specific decay rate modifiers of RothC dynamics (see Figure \ref{fig:alphas}) are not estimable using these simpler approaches.

The inferences made in this article, about soil-carbon cycling from the MTT dataset using CQUESST, provide valuable insights about soil-organic-carbon stocks and how agricultural practices can affect them.  Of particular note are the large carbon fluxes to the atmosphere using chemical fallow (PF) compared to other treatments, which is consistent with the results previously reported by \cite{Curtin2022}.  This lends strength to the observations made by other researchers \citep[e.g.,][]{Halvorson2002.1} that the use of crop-fallow rotations may lead to net carbon losses from the soil.  Furthermore, we observed that PP (i.e., permanent-pasture, no tillage) had a continual integration of plant organic carbon into the soil with large negative carbon fluxes (i.e., maximum sequestration of carbon).  Apart from PP and PF, we observed the highest posterior probabilities of soil-carbon sequestration for treatments In0, In1 and Im0, which employed one round of intensive tillage per year.  In our framework we observed a higher probability of sequestration with tillage, which may be due to smaller losses of surface residues and enhanced protection when organic matter is incorporated to depth.  Related to this, the work of \cite{cai2022} suggests that over time scales less than about 14 years, no-tillage agricultural treatments may show lower carbon stocks compared to those that employ conventional tillage, but that the difference diminishes beyond 14 years.

Our use of CQUESST to perform analysis of the designed MTT experiment, has allowed us to model and examine cropping-specific decay rates in the different carbon pools for a site in Lincoln, New Zealand. Within the presented framework that incorporates uncertainty in observations, process, and parameters, our results indicate that the decay rates of carbon pools vary as a function of management practices (i.e., tillage intensity and winter-cover-crop use), rather than being static rates consistent over all farms, as is typically assumed.  Treatment effects suggest that decay rates were higher in treatments without cover crops.

RothC dynamics relies on decay-rate modifiers (see $r_{i,t}$ in Equation (\ref{transitionEqStoch})) applied at a monthly time step to differentiate crop influences on soil moisture.  This allows RothC to slow decomposition in the presence of a crop to mimic the effect of drier soil on microbial activity.  It should be noted that in addition to soil-moisture dynamics, other potential mechanisms that are not captured in the RothC process model could occur.  It is now recognized that addition of fresh organic matter can positively prime the decomposition of native organic matter, with the impact diminishing over time \citep{Schiedung2023}.  In addition, elemental stoichiometry can control decay rates.  Nutrient availability, especially of nitrogen, is not modeled in RothC and therefore is not considered in CQUESST.  Similarly, differences in the DPM/RPM ratio ($r_{DPM/RPM}$; introduced in Section \ref{soilCarbonDynamics} and Supplemental Materials 2 and 4) for plant material entering the soil between summer crops and winter cover crops, are not represented in the deterministic mechanisms of RothC presented in Supplemental Materials 1 and 2.  Overall, field studies of soil carbon under cover cropping worldwide have been equivocal \citep{Poeplau2015, Chaplot2023}.  Our results suggest that the use of winter cover crops are sufficiently encouraging for soil-carbon sequestration, which warrants further investigation.

Analysis of the MTT using CQUESST involved a high-dimensional state-space model whose state vector was of length 252 (42 fields, each with 6 pools) and where dynamics evolved over 108 time steps (months).  Our implementation of CQUESST harnesses the computational power of {\tt Stan}, and it is noteworthy that the inclusion of the {\tt MapReduce} functionality in {\tt Stan} made our statistical analysis parallelizable and hence very efficient.  Of particular note is that {\tt Stan} allows for {\tt MapReduce} to be used in conjunction with {\tt MPI} \citep{mpi41} so that CQUESST can be made massively parallel on high-performance computing clusters.  Analyses therefore scale very well, allowing inference on very large numbers of field-plots without drastically increasing compute time beyond what it would be for a single field-plot.  For the MTT, CQUESST used 40 CPUs (10 CPUs per computing node) for each (parallel) Markov chain (70,000 samples per chain), with each of these taking between 33.6 and 36.6 hours to complete.  Overall, our analysis took roughly a day and a half and required 240 CPUs.  {\tt Stan} is a key enabler of analyses of the type presented here, where data have been collected at many independent sites but share some underlying parameters governing process dynamics.  This which would allow detailed analyses of carbon stocks (with uncertainty quantification) at broad geographical scales (e.g., for national soil-carbon accounting).

Understanding the complex biogeochemical cycles that take place in agricultural soils is critical to finding strategies to sequester carbon on the 37\% of Earth's surface that is used for agricultural production.  Of paramount importance are the collection of good quality, longitudinal datasets from field trials that can be used to gain insights into how carbon cycling differs in different climates, soils, tillage, and other practices.  This will be advantageous in mitigating climate change as well as improving the productivity of agricultural soils.  CQUESST provides a framework with Bayesian hierarchical modeling at its core to model complex soil-carbon dynamics in agricultural systems.  We envisage that this will enable scientists to inform and help their countries take steps towards net-zero carbon emissions by 2050.

\section*{Acknowledgements.}
Pagendam was supported by CSIRO's Digiscape Future Science Platform and Machine Learning and Artificial Intelligence Future Science Platform.  Cressie was supported by the Australian Research Council (ARC) Discovery Project DP190100180. Further, this material is based on work supported by the Air Force Office of Scientific Research under award number FA2386-23-1-4100.  The MTT data set was collected with scientific and technical support from Trish Fraser, Richard Gillespie, Weiwen Qiu, Chris Dunlop, Peg Gosden, and Sarah Glasson. The MTT trial was completed under the New Zealand Institute for Plant and Food Research Limited’s Sustainable Agro-ecosystems (SAE) programme, with funding from the Strategic Science Investment Fund (SSIF), contract C11X1702.  The authors gratefully acknowledge Josh Bowden for high-performance computing support.  Helpful comments from Gerard Heuvelink of the University of Wageningen are gratefully acknowledged, as are the editors' and reviewers' comments on an earlier submission.

\FloatBarrier
\vfill \null \pagebreak

\setcounter{equation}{0}
\renewcommand{\theequation}{S1.\arabic{equation}}
\setcounter{figure}{0}
\renewcommand{\thefigure}{S1.\arabic{figure}}
\setcounter{table}{0}
\renewcommand{\thetable}{S1.\arabic{table}}
\setcounter{section}{0}
\renewcommand{\thesection}{S1.\arabic{section}}

\section*{\LARGE{Supplemental Material 1: The Deterministic RothC Model.}}

In RothC \citep{Jenkinson1990}, the modeling of carbon cycling in soil is expressed mathematically according to the following set of deterministic (i.e., non-stochastic) equations, where we adopt notation that accommodates parameters varying from field-plot to field-plot (indexed by $i$) within the MTT.  For $i = 1, \dots, 42$,

\begin{align}\label{transitionEq}
D_{i, t + \Delta t} &= D_{i,t} e^{-r_{i, t + \Delta t} \frac{K_{D}}{12} \Delta t} + p_{P \rightarrow D} P_{i,t} + p_{M \rightarrow D} M_{i,t} \notag \\
R_{i, t + \Delta t} &= R_{i,t} e^{-r_{i, t + \Delta t} \frac{K_{R}}{12} \Delta t} + (1 - p_{P \rightarrow D}) P_{i,t} + p_{M \rightarrow R} M_{i,t}  \\
F_{i, t + \Delta t} &= F_{i,t} e^{-r_{i, t + \Delta t} \frac{K_{F}}{12} \Delta t} + p_{U \rightarrow F} U_{i,t} + p_{V  \rightarrow F}V_{i,t}  + p_{M \rightarrow F} M_{i,t} \notag \\
S_{i, t + \Delta t} &= S_{i,t} e^{-r_{i, t + \Delta t} \frac{K_{S}}{12} \Delta t} + p_{U  \rightarrow S} U_{i,t} + p_{V  \rightarrow S} V_{i,t} + p_{M \rightarrow S} M_{i,t}\notag \\
H_{i, t + \Delta t} &= H_{i,t} e^{-r_{i, t + \Delta t} \frac{K_{H}}{12} \Delta t} + p_{U  \rightarrow H} U_{i,t} + p_{V  \rightarrow H} V_{i,t}  + p_{M \rightarrow H} M_{i,t} \notag \\
I_{i, t + \Delta t} &= I_{i,t}, \notag
\end{align}

\noindent where

\begin{align}\label{processNodes}
U_{i,t} &\equiv \sum_{X \in \{D, R, F, S\}} \hspace{-5mm} X_{i, t}(1 - e^{-r_{i, t + \Delta t} \frac{K_{X}}{12}\Delta t}) \notag \\
V_{i,t} &\equiv H_{i,t} (1 - e^{-r_{i, t + \Delta t} \frac{K_{H}}{12}\Delta t}),
\end{align}

\noindent and where $\Delta t = 1$ month; $K_{D}$, $K_{R}$, $K_{F}$, $K_{S}$, and $K_{H}$ are respectively the annual decay-rates for carbon in the $D$, $R$, $F$, $S$, and $H$ pools for field-plot $i$; $r_{i, t}$ is a decay-rate modifier that is applied to the decay-rates as a result of changes in climate and ground cover; $U_{i,t}$ is the total carbon decayed from $D_{i,t}$, $R_{i,t}$, $F_{i,t}$, and $S_{i,t}$ pools in field-plot $i$, that remained in the soil from month $t$ to $t + 1$; $V_{i,t}$ is the carbon decayed from $H_{i,t}$ in field-plot $i$, that remained in the soil from month $t$ to $t + 1$; and $P_{i,t}$ is the addition of carbon from plant material to the soil in field-plot $i$ and month $t$. The role of each of the dynamic-process model parameters (e.g., $p_{P \rightarrow D}$, $p_{M \rightarrow D}$, etc.) is derived from soil-science considerations expanded in Supplemental Material 2.

\FloatBarrier
\vfill \null \pagebreak

\setcounter{equation}{0}
\renewcommand{\theequation}{S2.\arabic{equation}}
\setcounter{figure}{0}
\renewcommand{\thefigure}{S2.\arabic{figure}}
\setcounter{table}{0}
\renewcommand{\thetable}{S2.\arabic{table}}
\setcounter{section}{0}
\renewcommand{\thesection}{S2.\arabic{section}}


\section*{\LARGE{Supplemental Material 2: Dynamical Process Model Parameters.}}

The models defined in (\ref{transitionEqStoch}) of the main paper share two main classes of parameters: (i) decay rates, denoted by $K_X$ ($X \in \{ D, R, F, S, H \}$); and (ii) proportions of carbon routed between pools.  These are the deterministic parameters used in the deterministic model RothC.  For the former class, we generalize the decay rates (see (\ref{eq:alpha})) such that they are modeled as the product of $\kappa_X$ and $\alpha_{\tau}$, and the parameter model used for these is provided in Table \ref{priorTable1}.  For the parameters in the latter class, $p_{P \rightarrow D}$ describes the proportion of plant matter entering the D pool with the remainder entering the R pool; $p_{U \rightarrow F}$, $p_{U \rightarrow S}$, and $p_{U \rightarrow H}$ describe the proportions of decayed carbon $U_{i,t}$ that move into the $F$, $S$, and $H$ pools respectively; and $p_{V \rightarrow F}$, $p_{V \rightarrow S}$, and $p_{V \rightarrow H}$ describe the proportions of decayed carbon $V_{i,t}$ that moves into the $F$, $S$, and $H$ pools respectively.  The parameter model for this latter class is also given in Table \ref{priorTable1}.  We refer to the parameters $p_{X \rightarrow F}$, $p_{H \rightarrow S}$, $p_{clay}$, $\pi_{M \rightarrow Y}$, and $r_{DPM/RPM}$ as primary parameters that govern, respectively, the proportion of solid soil-carbon that enters pool $F$ from pool $X \in \{D, R, F, S\}$, the proportion of solid soil-carbon that enters pool $S$ from pool $H$, the proportion of clay in the soil, the proportion of manure entering pool $Y \in \{ D, R, F, S, H \}$, and the ratio of DPM to RPM in plant matter that enters the soil.   For the MTT site, $p_{clay}$ was estimated to be 0.16.  We also note that $r_{DPM/RPM}$ turns out to be a parameter that is discussed in Sections 4 and 5 of the main paper.

Several other parameters in (\ref{transitionEqStoch}) of the main paper are in fact \emph{derived} parameters from the set of primary parameters.  From these primary parameters, one obtains \citep{Jenkinson1977}:

\begin{align}
p_{P \rightarrow D} &= \frac{r_{DPM/RPM}}{1 + r_{DPM/RPM}} \notag \\
r_{CO2/Solid} &= 1.67(1.85 + 1.6e^{-7.86 p_{clay}}) \notag  \\
p_{U \rightarrow F} &= \frac{p_{X \rightarrow F}}{1 + r_{CO2/Solid}} \notag \\
p_{U \rightarrow S}  &= 0.0 \notag \\
p_{U \rightarrow H} &=  \frac{(1 - p_{X \rightarrow F})}{1 + r_{CO2/Solid}} \label{derivedParam1} \\
p_{V \rightarrow F} &= 0.0 \notag \\
p_{V \rightarrow S} &=  \frac{p_{H \rightarrow S}}{1 + r_{CO2/Solid}} \notag \\
p_{V \rightarrow H} &= \frac{1 - p_{H \rightarrow S}}{1 + r_{CO2/Solid}} \notag \\
p_{M \rightarrow Y} &= \frac{\pi_{M \rightarrow Y}}{\sum_{Q \in \{ D, R, F, S, H \}} \pi_{M \rightarrow Q} }. \notag
\end{align}

In equation (\ref{derivedParam1}) above, $r_{CO2/Solid}$ is the mass of CO$_2$ lost to the atmosphere for every unit of carbon mass that decomposes from a pool.  Decomposed carbon that is not lost from the soil as carbon dioxide is cycled to other carbon pools.  We note that  $r_{CO2/Solid}$ contains a number of empirically derived constants that model this parameter as a function of primary parameter $p_{clay}$, where the latter parameter is the proportion of the soil mass that can be considered clay material.  All of the parameters denoted as $p_{\cdot \rightarrow \cdot}$ are assumed to be identical across all field-plots and constant in time for the duration of the MTT.  To simplify dynamics, the parameters $p_{U \rightarrow S}$ and $p_{V \rightarrow F}$ are both put equal to zero in RothC v26.3, as do we in (\ref{transitionEqStoch}); these two biological pools are typically only very small relative to the total soil carbon.

\FloatBarrier
\vfill \null \pagebreak

\setcounter{equation}{0}
\renewcommand{\theequation}{S3.\arabic{equation}}
\setcounter{figure}{0}
\renewcommand{\thefigure}{S3.\arabic{figure}}
\setcounter{table}{0}
\renewcommand{\thetable}{S3.\arabic{table}}
\setcounter{section}{0}
\renewcommand{\thesection}{S3.\arabic{section}}


\section*{\LARGE{Supplemental Material 3: Geostatistical Diagnostics.}}

\section{Geostatistical Exploratory Data Analysis.}
\begin{figure} 
\begin{center} 
\includegraphics[width=0.9\textwidth]{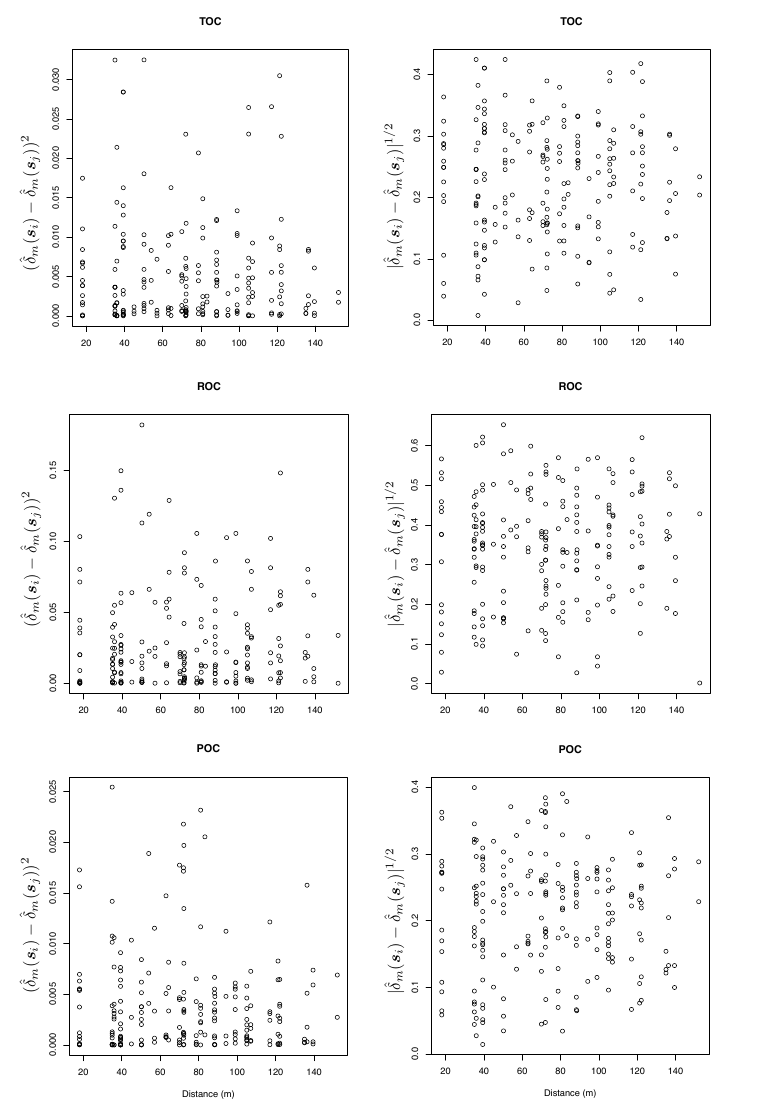}
\caption{Diagnostic plots to explore how the small-scale variation $\hat{\delta}_m(\boldsymbol{s})$ varies as a function of distance for measurements of TOC, POC, and ROC.\label{fig:spatialDep}}
\end{center}
\end{figure}

Here we explore whether the data from the Millennium Tillage Trial (MTT), soil-carbon measurements of TOC, POC and ROC exhibit spatial dependence.  Consider the spatial statistical model, 

$$ \log(Z_{m}(\boldsymbol{s})) = \mu_m(\boldsymbol{s}) + \delta_m(\boldsymbol{s}), $$

\noindent where $\boldsymbol{s}$ is the spatial coordinate for a sample of measurement  $m \in \{ \textup{TOC},$ $\textup{POC}, \textup{ROC} \}$, $\mu_m(\cdot)$ is a deterministic process representing the large-scale variation as a function of space, and $\delta_m(\boldsymbol{\cdot})$ is a stochastic process representing small-scale variation.  Since the MTT site was arranged as a matrix of plots, each sample location was defined as $\boldsymbol{s} = (x, y)^{\top}$, where $(0, 0)^{\top}$ is the center of the plot in the $(1,1)$ entry of the matrix, $x$ is the distance to the plot center in the row-wise direction, and $y$ is the distance to the plot centre in the column-wise direction.  The deterministic, large-scale variation was modeled via spatial trend as follows:

$$ \mu_m(\boldsymbol{s})  = \beta_{m, 0} + \beta_{m, r}x + \beta_{m, c}y + \beta_{m, rc}xy,$$

\noindent where $\beta_{m, 0}$, $\beta_{m, r}, \beta_{m, c}$, and $\beta_{m, rc}$ are regression parameters specific to measurement $m$.  To establish whether there was evidence of spatial covariance in the fine-scale stochastic process $\delta_m(\cdot)$, we first used ordinary least squares to estimate regression parameters, from which we defined an estimate of the (possibly spatially dependent) small-scale variation:

\begin{align*}
\hat{\delta}_m(\boldsymbol{s}) &= \log(Z_{m}(\boldsymbol{s})) - \hat{\beta}_{m, 0} + \hat{\beta}_{m, r}x + \hat{\beta}_{m, c}y + \hat{\beta}_{m, rc}xy,
\end{align*}

\noindent where $Z_m(\boldsymbol{s})$ is the measurement of type $m$, and $\hat{\beta}_{m, 0}$, $\hat{\beta}_{m, r}$, $\hat{\beta}_{m, c}$ and $\hat{\beta}_{m, rc}$ are the estimated regression parameters.  For each unique pair of locations $\boldsymbol{s}_i$, $\boldsymbol{s}_j$ that index the spatial dataset, $\mathcal{D}_m = \{ \hat{\delta}_m(\boldsymbol{s}_1), \dots, \hat{\delta}_m(\boldsymbol{s}_n) \}$, we computed $(\hat{\delta}_m(\boldsymbol{s}_i) - \hat{\delta}_m(\boldsymbol{s}_j))$ and the Euclidean distance $d_{i,j} = ||\boldsymbol{s}_i - \boldsymbol{s}_j||_2$.  Exploratory plots of $(\hat{\delta}_m(\boldsymbol{s}_i) - \hat{\delta}_m(\boldsymbol{s}_j))^2$  against $d_{i,j}$ and $|\hat{\delta}_m(\boldsymbol{s}_i) - \hat{\delta}_m(\boldsymbol{s}_j)|^{1/2}$ (see \cite{Cressie1993}, Section 2.4) were used to seek evidence of spatially structured dependence in the small-scale stochastic process $\delta_m(\cdot)$.  Those plots are given in Figure \ref{fig:spatialDep}, from which we saw no evidence of spatial structure.  Hence, we proceeded with the assumption that that $\text{cov}(\delta_m(\boldsymbol{s}_i) \delta_m(\boldsymbol{s}_j)) = 0, (\boldsymbol{s}_i \neq \boldsymbol{s}_j)$, for all three measurement types.

\begin{table}[h]\caption{Parameter model: prior probability distributions on terms in the Bayesian linear regression model for $\log(Z_{m}(\boldsymbol{s}))$.}\label{prior}
\begin{center}
\footnotesize
\begin{tabular}{  | p{1.3cm} | p{4.5cm} | p{5cm} | p{1.6cm} |}
\hline
Parameter & Description & Probability Distribution & Type  \\
\hline 
\hline 
$\beta_{\textup{m,0}}$ & Regression intercept parameter  & $\beta_{\textup{m,0}} \sim N(0, 1E6)$ & Uniformative \\
$\beta_{\textup{m,r}}$ & Regression slope parameter across field-plot rows & $\beta_{\textup{m,r}} \sim N(0, 1E6)$ & Uniformative \\
$\beta_{\textup{m,c}}$ & Regression slope parameter across field-plot columns  & $\beta_{\textup{m,c}} \sim N(0, 1E6)$ & Uniformative \\
$\beta_{\textup{m,rc}}$ & Regression interaction parameter across field-plot rows and columns  & $\beta_{\textup{m,rc}} \sim N(0, 1E6)$ & Uniformative \\
$\sigma^2_{\textup{m}}$ & Residual error variance & $\sigma^2_{\textup{m}} \sim \textup{Inverse Gamma}(0.001, 0.001)$ & Uniformative \\
\hline
\end{tabular}
\end{center}
\end{table}

\section{Estimation of Measurement Errors in Observed Soil-Carbon Measurable Fractions.}

With no evidence of spatial structure in the small scale stochastic process, we fitted the model, 

$$ \log(Z_{m}(\boldsymbol{s})) = \mu_m(\boldsymbol{s}) + \delta_m(\boldsymbol{s}) \quad (\delta_m(\boldsymbol{s}) \sim N(0, \sigma^2_m)), $$

\noindent as a Bayesian linear model with independent normal and inverse-gamma conjugate prior distributions, respectively on the parameter vector $\boldsymbol{\theta} = (\beta_{m,0}, \beta_{m,r}, \beta_{m,c}, \beta_{m,rc}, \sigma^2_m)^{\top}$.  These prior distributions are provided in Table \ref{prior}.  The estimate of $\sigma^2_m$ obtained for each measurement type, were used to devise informative prior distributions when using the CQUESST framework to analyze the MTT data.

\FloatBarrier
\vfill \null \pagebreak

\setcounter{equation}{0}
\renewcommand{\theequation}{S4.\arabic{equation}}
\setcounter{figure}{0}
\renewcommand{\thefigure}{S4.\arabic{figure}}
\setcounter{table}{0}
\renewcommand{\thetable}{S4.\arabic{table}}
\setcounter{section}{0}
\renewcommand{\thesection}{S4.\arabic{section}}


\section*{\LARGE{Supplemental Material 4: Parameter Model.}}

Here we outline the parameter model used in applying CQUESST to the Millennium Tillage Trial.  Prior distributions over the parameters of the model are provided in Tables \ref{priorTable1}, \ref{priorTable2}, \ref{priorTableIDataModel} and \ref{priorTableIC}.  Specifically, Table \ref{priorTable1} gives the priors for the soil-carbon-cycling parameters in (\ref{transitionEqStoch}); Table \ref{priorTable2} gives the priors for process-error variance parameters; Table \ref{priorTableIDataModel} gives the priors on error variances for measured carbon fractions in (\ref{datamodel}); and Table \ref{priorTableIC} gives prior distributions for initial conditions of the soil-carbon pools defined in (\ref{transitionEqStoch}).

In these tables, $TN_a^b$ denotes a truncated normal distribution that has been truncated on the left at $a$ and on the right at $b$.  We use truncated normal priors because all of the parameters in our model are bounded either from below or are within the interval $[0, 1]$, and because they provide a convenient and easily interpretable way to incorporate prior information.  Other choices for priors on scale parameters in hierarchical models have been advocated \citep[e.g.,][]{Gelman2006, Polson2012}.  Where $\textup{Inverse Gamma}(\alpha, \beta)$ priors were used, $\alpha$ denotes the shape parameter and $\beta$ the scale parameter.

Finally, $\sigma^2_{D}$, $\sigma^2_{R}$, $\sigma^2_{F}$, $\sigma^2_{S}$, and $\sigma^2_{H}$ are not parameters that exist in the deterministic RothC model, and they are not biophysical parameters that have been estimated in past studies.  Priors on these parameters were specified using inverse-Gamma distributions with shape and scale parameters chosen so that $\textup{Pr}(e^{\eta_X} < 0.9) \approx 0.01$ and $\textup{Pr}(e^{\eta_X} > 1.1) \approx 0.01$, where for $X \in \{ D, R, F, S, H \}$, $\eta_X$ represents the process error distributed as $\eta_X \sim N(-\tfrac{\sigma^2_X}{2}, \sigma^2_X)$.  This results in the inverse-Gamma's shape parameter being set at 403.4 and its scale parameter at 0.318 (see Table \ref{priorTable2}).  Since $\eta_X$ is additive on the log-scale, $e^{\eta_X}$ represents the corresponding multiplicative process error on the natural scale where the soil carbon is cycling.  The values $0.9$ and $1.1$ were chosen so that the majority of the prior probability mass for $\sigma^2_{X}$ generated dynamics within $\pm 10\%$ of RothC's deterministic evolution of the soil carbon.

\begin{table}\caption{Parameter model: prior probability distributions on soil-carbon-cycling parameters in (\ref{transitionEqStoch}) and primary parameters.}\label{priorTable1}
\begin{center}
\footnotesize
\begin{tabular}{ | p{1.5cm} | p{5.0cm} | p{4.5cm} | p{1.8cm} |}
\hline
Parameter & Description & Probability Distribution & Type \\
\hline 
\hline 
$\kappa_{D}$ & Decomposition rate constant (y$^{-1}$) for $D$ (decomposable plant material).  & $\kappa_{D} \sim TN_{5.0}^{20.0}(10.0, (0.5)^2)$ & Informative \\
$\kappa_{R}$ &  Decomposition rate constant (y$^{-1}$) for $R$ (resistant plant material).   & $\kappa_{R} \sim TN_{0.05}^{5.0}(0.07, (0.0035)^2)$ & Informative \\
$\kappa_{F}$ &  Decomposition rate constant (y$^{-1}$) for $F$ (fast microbial biomass).   & $\kappa_{F} \sim TN_{0.3}^{1.0}(0.66, (0.033)^2)$ & Informative \\
$\kappa_{S}$ &  Decomposition rate constant (y$^{-1}$) for $S$ (slow microbial biomass).   & $\kappa_{S} \sim TN_{0.3}^{1.0}(0.66, (0.033)^2)$ & Informative \\
$\kappa_{H}$ &  Decomposition rate constant (y$^{-1}$) for $H$ (humus).   & $\kappa_{H} \sim TN_{0.005}^{0.05}(0.02, (0.001)^2)$  & Informative \\
$\alpha_{\tau}$ ($\tau \in \mathcal{T}$) & Multiplicative tillage-cropping treatment effect applied to decomposition rates. & $\log(\alpha_{\tau}) \sim TN_{-5}^{5}(0.0, 1.0)$ & Weakly Informative \\
$\pi_{M \rightarrow D}$ & Proportion of manure to D pool & $\pi_{M \rightarrow D} \sim TN_0^{1}(0.49, (0.01)^2)$ & Informative \\
$\pi_{M \rightarrow R}$ & Proportion of manure to R pool & $\pi_{M \rightarrow R} \sim TN_0^{1}(0.49, (0.01)^2)$ & Informative \\
$\pi_{M \rightarrow F}$ & Proportion of manure to F pool & $\pi_{M \rightarrow F} \sim TN_0^{1}(0.0, (0.01)^2)$ & Informative \\
$\pi_{M \rightarrow S}$ & Proportion of manure to S pool & $\pi_{M \rightarrow S} \sim TN_0^{1}(0.0, (0.01)^2)$ & Informative \\
$\pi_{M \rightarrow H}$ & Proportion of manure to H pool & $\pi_{M \rightarrow H} \sim TN_0^{1}(0.02, (0.01)^2)$ & Informative \\
$p_{X \rightarrow F}$ & Proportion of soil-carbon from $X \in \{D, R, F, S \}$ to $F$. & $TN_0^{1}(0.46, (0.01)^2)$ & Informative \\
$p_{H \rightarrow S}$ & Proportion of soil-carbon from $H$ to $S$. & $p_{H \rightarrow S} \sim TN_0^{1}(0.46, (0.01)^2)$ & Informative \\
$p_{clay}$ & Proportion of the soil that is clay & $p_{clay} \sim TN_0^{1}(0.16, (0.02)^2)$ & Informative \\
$r_{DPM/RPM}$ & Ratio of decomposable to resistant carbon in plant material. & $r_{DPM/RPM} \sim TN_0^{\infty}(1.44, (0.5)^2)$  & Informative \\
\hline
\end{tabular}
\end{center}
\end{table}

\begin{table}\caption{Parameter model: prior probability distributions on process-error variance parameters in (\ref{transitionEqStoch}).}\label{priorTable2}
\begin{center}
\footnotesize
\begin{tabular}{ | p{1.5cm} | p{5.0cm} | p{4.5cm} | p{1.8cm} |}
\hline
Parameter & Description & Probability Distribution & Type \\
\hline 
\hline 
$\sigma^2_D$ &  Variance of additive process noise of $D$ pool.  & $\textup{Inverse Gamma}(403.4, 0.318)$  & Informative \\
$\sigma^2_R$ &  Variance of additive process noise of $R$ pool.  & $\textup{Inverse Gamma.}(403.4, 0.318)$  & Informative \\
$\sigma^2_F$ &  Variance of additive process noise of $F$ pool.  & $\textup{Inverse Gamma}(403.4, 0.318)$ & Informative \\
$\sigma^2_S$ &  Variance of additive process noise of $S$ pool.  & $\textup{Inverse Gamma}(403.4, 0.318)$ & Informative \\
$\sigma^2_H$ &  Variance of additive process noise of $H$ pool.  & $\textup{Inverse Gamma}(403.4, 0.318)$ &  Informative \\
\hline
\end{tabular}
\end{center}
\end{table}

\begin{table}\caption{Parameter model: prior probability distributions on error variances for measured carbon fractions.}\label{priorTableIDataModel}
\begin{center}
\footnotesize
\begin{tabular}{ | p{1.5cm} | p{5.0cm} | p{4.5cm} | p{1.8cm} |}
\hline
Parameter & Description & Probability Distribution & Type  \\
\hline 
\hline 
$\sigma^2_{\textup{POC}}$ &  Measurement error variance for log(POC).  & $\text{Inverse Gamma}(\frac{21}{2}, 0.039)$ & Informative \\
$\sigma^2_{\textup{ROC}}$ &  Measurement error variance for  log(ROC). & $\text{Inverse Gamma.}(\frac{21}{2}, 0.290)$ & Informative \\
$\sigma^2_{\textup{TOC}}$ &  Measurement error variance for log(TOC). & $\text{Inverse Gamma}(\frac{21}{2}, 0.053)$ & Informative \\
\hline
\end{tabular}
\end{center}
\end{table}

\begin{table}\caption{Parameter model: prior probability distributions on initial conditions of the soil-carbon pools defined in (\ref{transitionEqStoch}).}\label{priorTableIC}
\begin{center}
\footnotesize
\begin{tabular}{ | p{1.5cm} | p{5.0cm} | p{4.5cm} | p{1.8cm} |}
\hline
Initial Condition & Description & Probability Distribution & Type \\
\hline 
\hline 
$D_{i, 0}$ &  The initial state of $D$ in field-plot $i$ (Mg/ha).  & $TN_{0}^{\infty}(0.0, (0.1)^2)$ & Informative \\
$R_{i, 0}$ &  The initial state of $R$ in field-plot $i$ (Mg/ha). & $TN_{0}^{\infty}(0.0, (100.0)^2)$  & Uninformative \\
$F_{i, 0}$ &  The initial state of $F$ in field-plot $i$  (Mg/ha). & $TN_{0}^{\infty}(0.0, (0.01)^2)$ & Informative \\
$S_{i, 0}$ &  The initial state of $S$ in field-plot $i$ (Mg/ha). & $TN_{0}^{\infty}(0.0, (0.01)^2)$ & Informative \\
$H_{i, 0}$ &  The initial state of $H$ in field-plot $i$ (Mg/ha). & $TN_{0}^{\infty}(0.0, (100.0)^2)$ &  Uninformative\\
$I_{i, 0}$ &  The initial state of $D$ in field-plot $i$ (Mg/ha).  & $TN_{0}^{\infty}(0.0, (10.0)^2)$ &  Uninformative \\
\hline
\end{tabular}
\end{center}
\end{table}

\FloatBarrier
\vfill \null \pagebreak

\setcounter{equation}{0}
\renewcommand{\theequation}{S5.\arabic{equation}}
\setcounter{figure}{0}
\renewcommand{\thefigure}{S5.\arabic{figure}}
\setcounter{table}{0}
\renewcommand{\thetable}{S5.\arabic{table}}
\setcounter{section}{0}
\renewcommand{\thesection}{S5.\arabic{section}}

\section*{\LARGE{Supplemental Material 5: Sensitivity Analysis of CQUESST.}}

\section{Sensitivity-Analysis Scenarios.}
We undertook a sensitivity analysis to explore how posterior distributions and conclusions drawn from them were affected by changes in the prior distributions used.  The sensitivity analysis considered changes to two sets of parameters, the first being the soil carbon decay rates $\kappa_X$, and the second being the process-error variances, $\sigma^2_X$, for each of the latent soil-carbon pools $X \in \{ D, R, F, S, H\}$.  The first group of parameters are important parameters that have important biogeochemical meaning in the RothC process model, whereas the second set are important statistical parameters that control additive uncertainty in the process dynamics not otherwise captured by the deterministic RothC model.

We denote three sets of prior distributions used with CQUESST as scenario N, A, and B.  Scenario N represents no change to the CQUESST model used in the main paper, with prior distributions as defined in Supplemental Material 3.  Scenario A represents fitting the model with more diffuse priors on the decay rates, $\kappa_X$, for each of the latent soil-carbon pools.  The priors for the decay rate parameters in scenario A were altered from those of scenario N so that the location parameters remained unchanged, but the variances were inflated by a factor of 4 as per Table \ref{sensitivityA_Table1}.  For scenario B, the priors for the $\sigma^2_X$ were altered from those of scenario N so that the mean of the prior remained the same, but the variance was inflated by a factor of 4 as detailed in Table \ref{sensitivityB_Table1}.

For each of the three scenarios (N, A, and B), we reran the Markov Chain Monte Carlo of CQUESST and obtained samples from six independent Markov chains, using 20,000 samples for warm-up that were subsequently discarded, and then 50,000 samples for posterior inference in each of these.  For more precise inferences, these were further thinned by taking every tenth sample, resulting in 5,000 approximately independent samples for each of the six chains.  As in the main paper, diagnostics of the chains were undertaken as per Supplemental Material 4 with all scenarios demonstrating strong evidence of converging to the posterior distribution ($\hat{R} < 1.01$ for all parameters).

\begin{table}[h]\caption{Altered prior probability distributions for soil-carbon decay rates used in scenario A.  These are less informative than those used in scenario N.}\label{sensitivityA_Table1}
\begin{center}
\footnotesize
\begin{tabular}{ | p{1.5cm} | p{5.0cm} | p{4.5cm} | p{1.8cm} |}
\hline
Parameter & Description & Probability Distribution & Type \\
\hline 
\hline 
$\kappa_{D}$ & Decomposition rate constant (y$^{-1}$) for $D$ (decomposable plant material).  & $\kappa_{D} \sim TN_{5.0}^{20.0}(10.0, (1.0)^2)$ & Less Informative \\
$\kappa_{R}$ &  Decomposition rate constant (y$^{-1}$) for $R$ (resistant plant material).   & $\kappa_{R} \sim TN_{0.05}^{5.0}(0.07, (0.007)^2)$ & Less Informative \\
$\kappa_{F}$ &  Decomposition rate constant (y$^{-1}$) for $F$ (fast microbial biomass).   & $\kappa_{F} \sim TN_{0.3}^{1.0}(0.66, (0.066)^2)$ & Less Informative \\
$\kappa_{S}$ &  Decomposition rate constant (y$^{-1}$) for $S$ (slow microbial biomass).   & $\kappa_{S} \sim TN_{0.3}^{1.0}(0.66, (0.066)^2)$ & Less Informative \\
$\kappa_{H}$ &  Decomposition rate constant (y$^{-1}$) for $H$ (humus).   & $\kappa_{H} \sim TN_{0.005}^{0.05}(0.02, (0.002)^2)$  & Less Informative \\
\hline
\end{tabular}
\end{center}
\end{table}

\begin{table}\caption{Altered prior probability distributions for process-error variances used in scenario B.  These are less informative than those used in scenario N.}\label{sensitivityB_Table1}
\begin{center}
\footnotesize
\begin{tabular}{ | p{1.5cm} | p{5.0cm} | p{4.0cm} | p{2.5 cm} |}
\hline
Parameter & Description & Probability Distribution & Type \\
\hline 
\hline 
$\sigma^2_D$ &  Variance of additive process noise of $D$ pool.  & $\textup{Inverse Gamma}(102.4, 0.08)$  & Less Informative \\
$\sigma^2_R$ &  Variance of additive process noise of $R$ pool.  & $\textup{Inverse Gamma}(102.4, 0.08)$  & Less Informative \\
$\sigma^2_F$ &  Variance of additive process noise of $F$ pool.  & $\textup{Inverse Gamma}(102.4, 0.08)$ & Less Informative \\
$\sigma^2_S$ &  Variance of additive process noise of $S$ pool.  & $\textup{Inverse Gamma}(102.4, 0.08)$ & Less Informative \\
$\sigma^2_H$ &  Variance of additive process noise of $H$ pool.  & $\textup{Inverse Gamma}(102.4, 0.08)$ &  Less Informative \\
\hline
\end{tabular}
\end{center}
\end{table}

\section{Sensitivity Analysis Results.}
\begin{figure} 
\begin{center}
\includegraphics[width=0.75\textwidth]{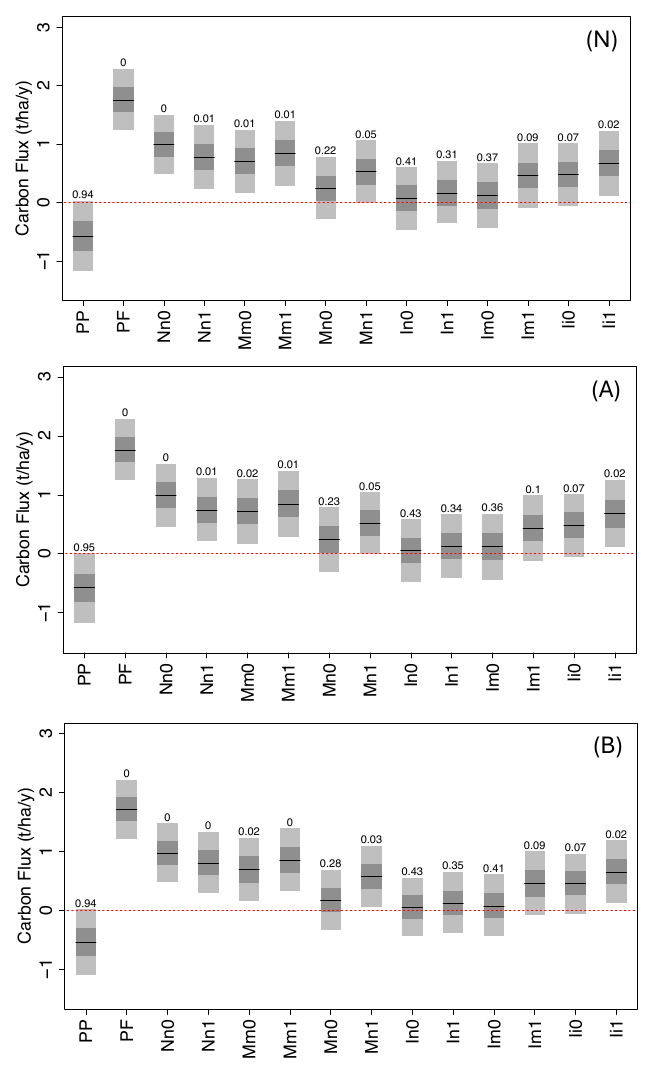}
\caption{Posterior predictive distributions of soil-carbon fluxes ($A^{(\tau)}$) over the duration of the Millennium Tillage Trial (MTT) for the different treatments $\tau$.  Plots are shown for scenarios N (no change to the priors used in the main paper), A (more diffuse priors on soil-carbon decay rates), and B (more diffuse priors on process-error variances).}
\label{Sensitivity_Flux}
\end{center} 
\end{figure}

\begin{figure} 
\begin{center}
\includegraphics[width=0.75\textwidth]{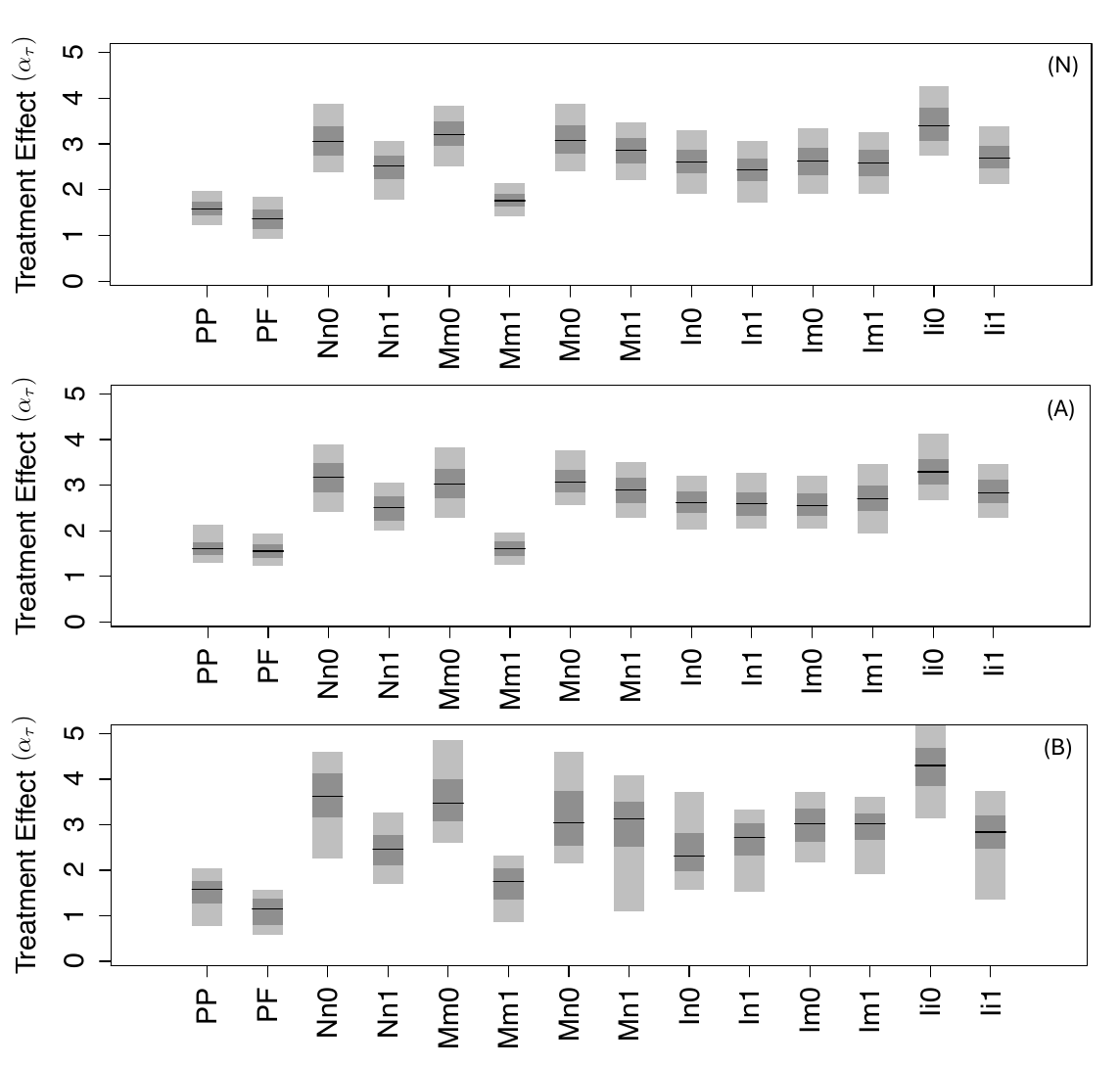}
\caption{Posterior predictive distributions of the Millennium Tillage Trial (MTT) treatment effects ($\alpha_{\tau}$) for different treatments $\tau$.  Plots are shown for scenarios N (no change to the priors used in the main paper), A (more diffuse priors on soil-carbon decay rates), and B (more diffuse priors on process error variances).}
\label{Sensitivity_Alpha}
\end{center} 
\end{figure}

\begin{figure} 
\begin{center}
\includegraphics[width=1.0\textwidth]{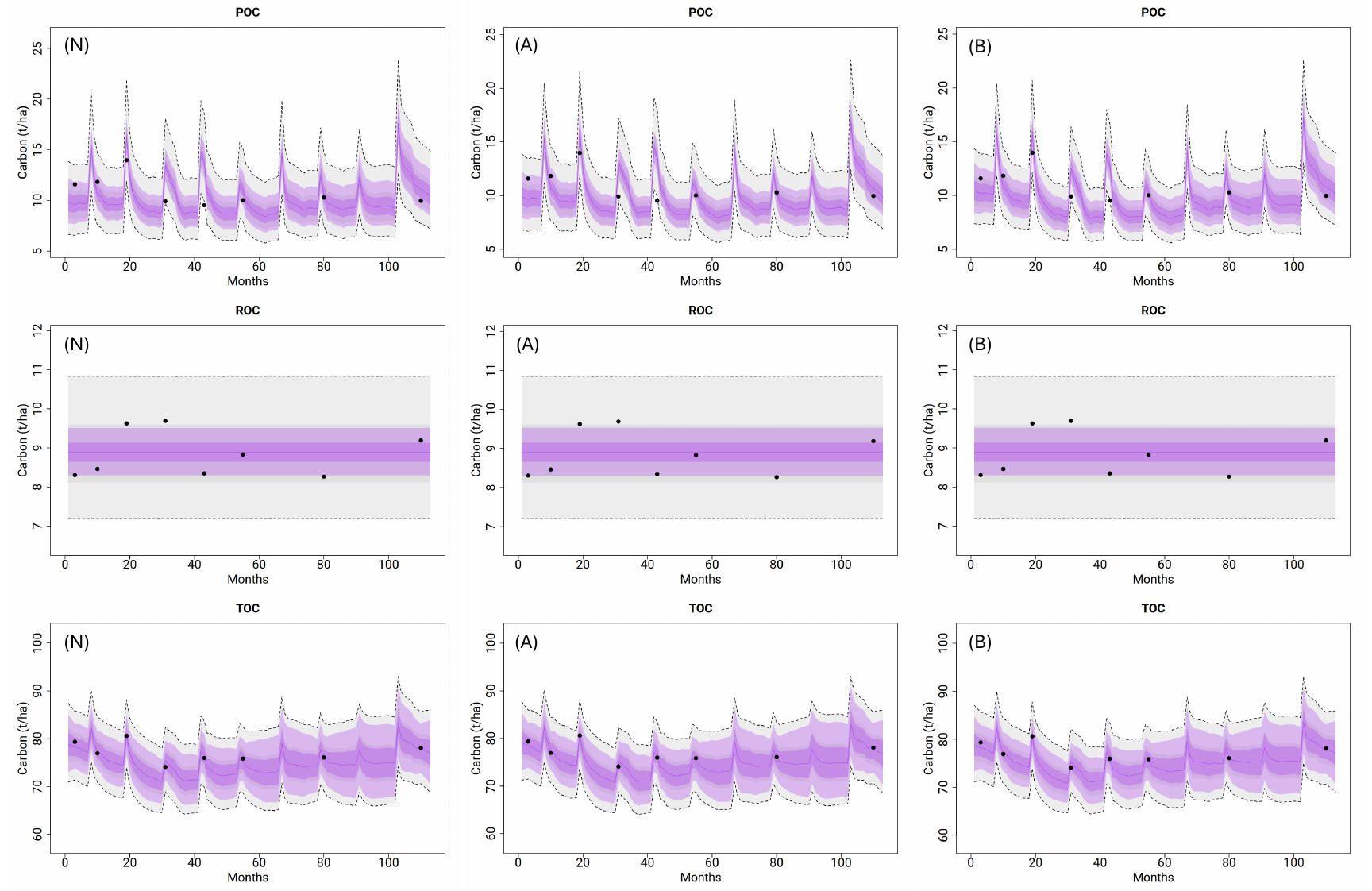}
\caption{Posterior predictive distributions of the combined latent soil-carbon pools for each observation type (observations shown as black dots) for the second field-plot with tillage treatment Mn0.  Plots are shown for scenarios N (no change to the priors used in the main paper), A (more diffuse priors on soil-carbon decay rates), and B (more diffuse priors on process-error variances).}
\label{Sensitivity_Trajectories}
\end{center} 
\end{figure}

To assess the affects of perturbing the priors on the soil-carbon decay rates and the measurement-error variances, we computed posterior distributions on important quantities that were presented in the main paper.  Of particular interest were the carbon fluxes over the duration of the Millennium Tillage Trial (MTT) and the treatment effects ($\alpha_{\tau}$).  We present the same type of plots presented in the main paper for scenarios N, A and B in Figures \ref{Sensitivity_Flux} and \ref{Sensitivity_Alpha}.  We also recreated Figure \ref{fig:Observed_MN0_Field2} from the main paper for each of the three scenarios to assess whether there were obvious differences in how the latent soil carbon pools fit observed data.  These plots of the sampled trajectories are presented in Figure \ref{Sensitivity_Trajectories}.

\begin{figure} 
\begin{center}
\includegraphics[width=1\textwidth]{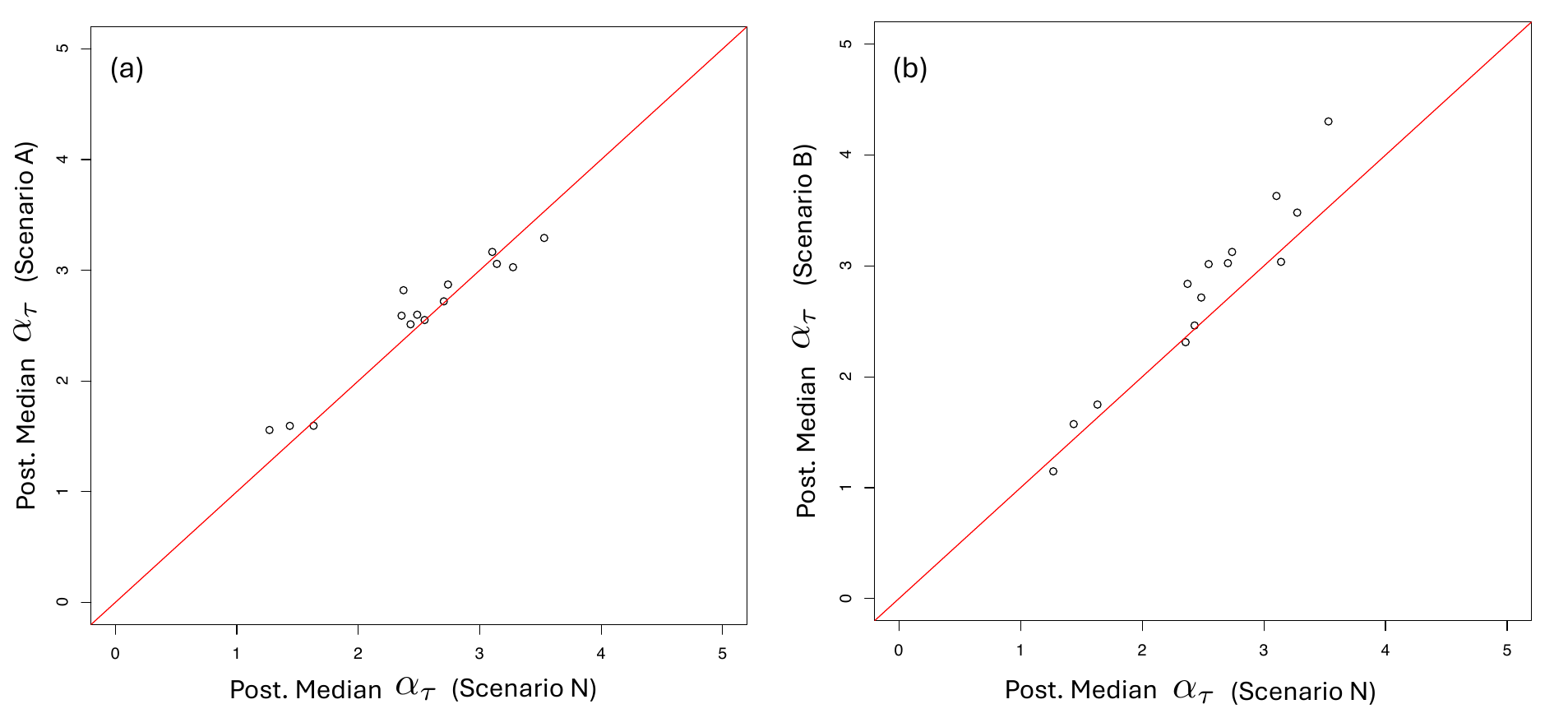}
\caption{Scatter plots of posterior medians of $\alpha_{\tau}$ under different sensitivity-analysis scenarios with the red line showing perfect 1:1 correspondence.  Subplot (a) shows scenario A (more diffuse priors on soil-carbon decay rates) versus scenario N (no change to the priors used in the main paper).  Subplot (b) shows scenario B (more diffuse priors on process-error variances) versus scenario N (no change to the priors used in the main paper).}
\label{Sensitivity_Alpha_1to1}
\end{center} 
\end{figure}

\begin{figure} 
\begin{center}
\includegraphics[width=1\textwidth]{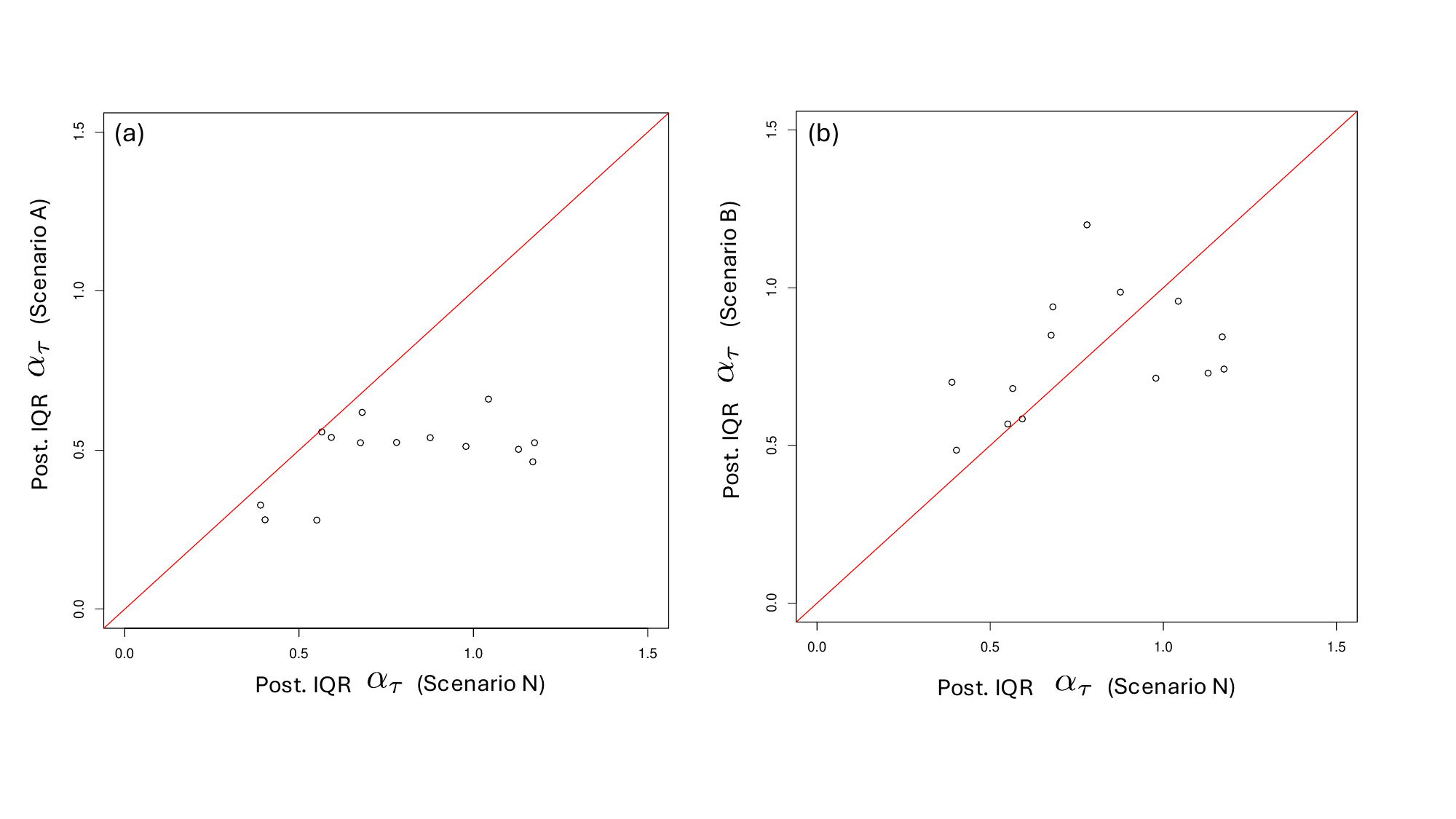}
\caption{Scatter plots of posterior interquartile range (IQR) of $\alpha_{\tau}$ under different sensitivity-analysis scenarios with the red line showing perfect 1:1 correspondence.  Subplot (a) shows scenario A (more diffuse priors on soil-carbon decay rates) versus scenario N (no change to the priors used in the main paper).  Subplot (b) shows scenario B (more diffuse priors on process-error variances) versus scenario N (no change to the priors used in the main paper).}
\label{Sensitivity_Alpha_1to1_IQR}
\end{center} 
\end{figure}

\begin{figure} 
\begin{center}
\includegraphics[width=1\textwidth]{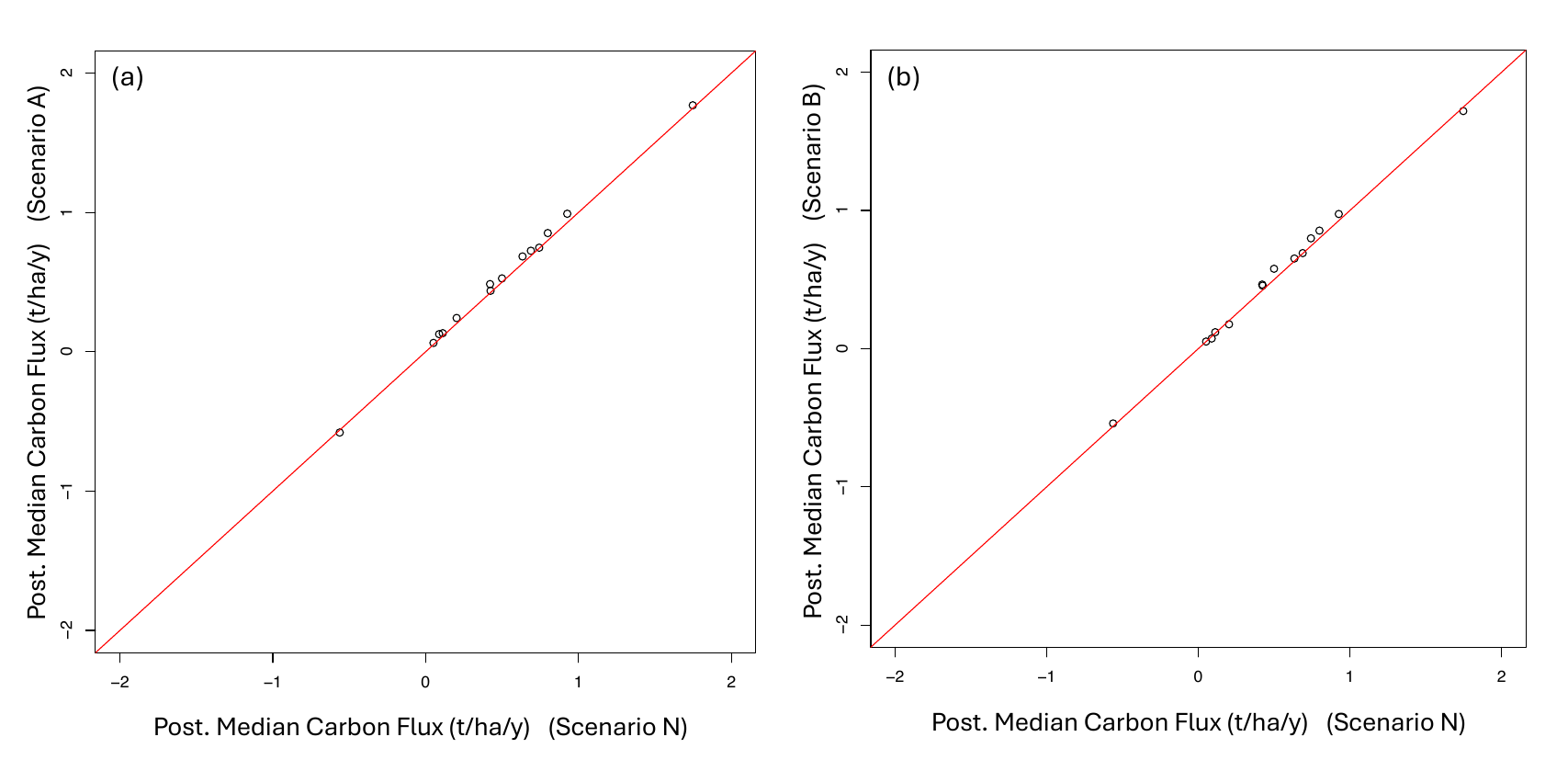}
\caption{Scatter plots of posterior medians of carbon flux (estimated for each of the different treatments) under different sensitivity-analysis scenarios with the red line showing perfect 1:1 correspondence.  Subplot (a) plots scenario A (more diffuse priors on soil-carbon decay rates) versus scenario N (no change to the priors used in the main paper).  Subplot (b) plots scenario B (more diffuse priors on process-error variances) versus scenario N (no change to the priors used in the main paper).}
\label{Sensitivity_Flux_1to1}
\end{center} 
\end{figure}

\begin{figure} 
\begin{center}
\includegraphics[width=1\textwidth]{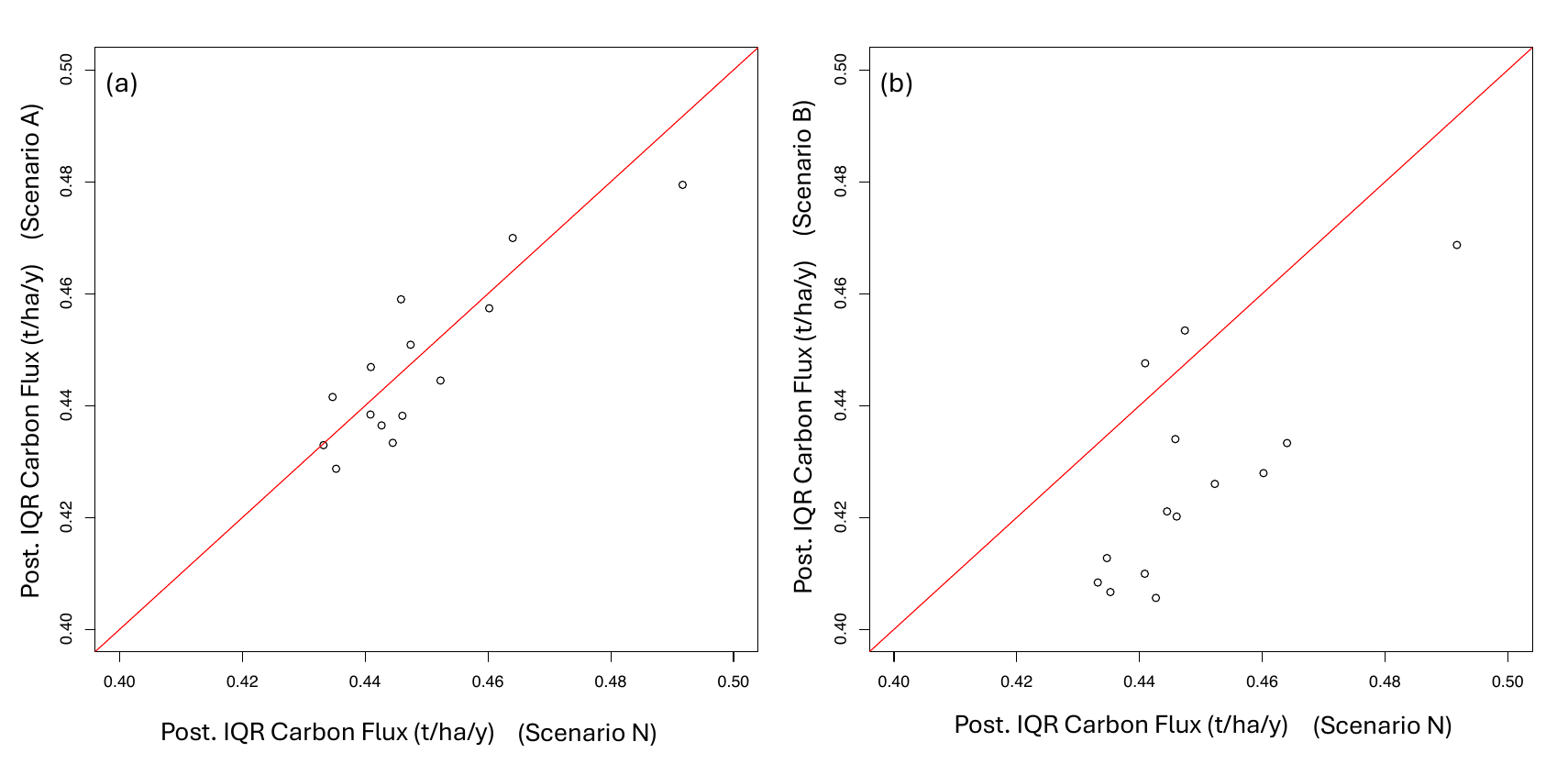}
\caption{Scatter plot of posterior interquartile range (IQR) of carbon flux (estimated for each of the different treatments) under different sensitivity analysis scenarios with the red line showing perfect 1:1 correspondence.  Subplot (a) plots scenario A (more diffuse priors on soil-carbon decay rates) scenario N (no change to the priors used in the main paper).  Subplot (b) plots scenario B (more diffuse priors on process-error variances) versus scenario N (no change to the priors used in the main paper).}
\label{Sensitivity_Flux_1to1_IQR}
\end{center} 
\end{figure}

Overall, when comparing scenarios N, A, and B, there is very little difference in the results we obtained.  This suggests that our results were relatively robust to the prior distributions chosen.  Of particular note are that the posterior medians are nearly identical for all three scenarios.  We also see great similarity between the trajectories for the three observation types plotted in Figure \ref{Sensitivity_Trajectories}, which suggests that perturbing the prior distributions had negligible effect on the estimated soil carbon dynamics in field-plots.  Plots of the treatment effects ($\alpha_{\tau}$) for the three scenarios showed little difference between scenario N and A, with similar but slightly more diffuse posteriors over the $\alpha_{\tau}$ in scenario B.  To aid comparison of the scenarios, in Figure \ref{Sensitivity_Alpha_1to1} we plot the posterior medians of $\alpha_{\tau}$ for scenario A versus those for N (subplot (a)), and for scenario B versus N (subplot (b)).  Similarly, in Figure \ref{Sensitivity_Flux_1to1} we plot the posterior median carbon fluxes for the different treatments for scenarios A and B versus scenario N.  For the posterior median fluxes, almost no change is seen in our point estimates when the priors were made more diffuse. These scatter plots demonstrate clearly that the central tendencies of the posterior distributions were largely unchanged for the more diffuse priors.  In Figures \ref{Sensitivity_Alpha_1to1_IQR} and \ref{Sensitivity_Flux_1to1_IQR} we also plot the posterior interquartile ranges (IQR) for $\alpha_{\tau}$ and carbon flux for the MTT treatments.  For $\alpha_{\tau}$, the posterior IQR tended to be higher for scenario N than for scenario A.  For scenarios scenarios N and B, the IQRs were similar.  For carbon flux, the posterior IQRs did not vary greatly between scenario N, A, and B, but there was a tendency for the posterior IQR to be slightly lower for scenario B than they were in scenario N.

\FloatBarrier
\vfill \null \pagebreak

\setcounter{equation}{0}
\renewcommand{\theequation}{S6.\arabic{equation}}
\setcounter{figure}{0}
\renewcommand{\thefigure}{S6.\arabic{figure}}
\setcounter{table}{0}
\renewcommand{\thetable}{S6.\arabic{table}}
\setcounter{section}{0}
\renewcommand{\thesection}{S6.\arabic{section}}

\section*{\LARGE{Supplemental Material 6: MCMC Diagnostics.}}

When performing Bayesian inference with an MCMC algorithm, it is important to verify that the Markov chain provides a representative set of samples from the posterior distribution $p(\mathbf{Y}, \boldsymbol{\theta} | \mathbf{Z})$.  A common metric used to assess convergence is the $\hat{R}$ statistic \citep{Gelman2004.1}.  This metric assesses the convergence of each element $\gamma_k$ in the vector of all sampled random variables, $\boldsymbol{\gamma} = (\boldsymbol{\theta}^{\top}, \mathbf{Y}^{\top})^{\top}$,  using $m$ independent Markov chains, each containing $n$ samples.  For each $\gamma_k$, the statistic is calculated from the ratio of two variance estimators, namely:

\begin{equation*}
\hat{R}_{\gamma_k} = \sqrt{\frac{\widehat{\textup{Var}}^{+}(\gamma_k | \mathbf{Z})}{W}},
\end{equation*}

\noindent where  $\gamma_{i,j,k}$ denotes the $i$th sample from one of $j = 1, \dots, m$ Markov chains, each consisting of $n$ samples.

\begin{align*}
& \widehat{\textup{Var}}^{+}(\gamma_k | \mathbf{Z}) = \frac{n-1}{n}W + \frac{1}{n}B, \\
& B = \frac{n}{m - 1}\sum_{j = 1}^m (\bar{\gamma}_{\cdot j k} - \bar{\gamma}_{\cdot \cdot k})^2, \quad \bar{\gamma}_{\cdot j k} = \frac{1}{n} \sum_{i = 1}^n {\gamma}_{i j k}, \quad \bar{\gamma}_{\cdot \cdot k} = \frac{1}{nm} \sum_{i = 1}^n \sum_{j = 1}^m {\gamma}_{i j k},  \\
& W = \frac{1}{m} \sum_{j = 1}^m s_{jk}^2, \quad s_{jk}^2 = \frac{1}{n-1}\sum_{i = 1}^n (\bar{\gamma}_{ijk} - \bar{\gamma}_{\cdot jk})^2.
\end{align*}

\noindent  When $\hat{R} < 1.01$ for every $\gamma_k \in \boldsymbol{\gamma}$, the aggregated set of samples from the $m$ chains is considered to provide a set of samples from which reliable posterior inferences can be made.  

In total, we obtained samples from $m = 6$ independent Markov chains, using 20,000 samples for warm-up that were subsequently discarded, and then 50,000 samples for posterior inference in each of these.  For more precise inferences, these were further thinned by taking every tenth sample, resulting in $n = 5,000$ approximately independent samples for each of the six chains. For all $\gamma_k$, we computed $\hat{R}$ and assessed that posteriors were reliable (see Tables \ref{convergenceTable1} and \ref{convergenceTable2}, \ref{convergenceTable3}).

\begin{table}\caption{MCMC convergence diagnostics for soil-carbon-cycling parameters in (\ref{transitionEqStoch}) and primary parameters.}\label{convergenceTable1}
\begin{center}
\footnotesize
\begin{tabular}{ | p{1.5cm} | p{8.0cm} | p{2cm} | }
\hline
Parameter & Description & $\hat{R}$ \\
\hline 
\hline 
$\kappa_{D}$ & Decomposition rate constant (y$^{-1}$) for $D$ (decomposable plant material).  & 1.000  \\
$\kappa_{R}$ &  Decomposition rate constant (y$^{-1}$) for $R$ (resistant plant material).   & 1.000 \\
$\kappa_{F}$ &  Decomposition rate constant (y$^{-1}$) for $F$ (fast microbial biomass).   & 1.000 \\
$\kappa_{S}$ &  Decomposition rate constant (y$^{-1}$) for $S$ (slow microbial biomass).   & 0.999 \\
$\kappa_{H}$ &  Decomposition rate constant (y$^{-1}$) for $H$ (humus).   & 1.000 \\
$\alpha_{PP}$ & Multiplicative tillage-cropping treatment effect for treatment PP. & 1.000  \\
$\alpha_{Nn0}$ & Multiplicative tillage-cropping treatment effect for treatment Nn0. &1.000  \\
$\alpha_{Nn1}$ & Multiplicative tillage-cropping treatment effect for treatment Nn1. & 1.000  \\
$\alpha_{Mm0}$ & Multiplicative tillage-cropping treatment effect for treatment Mm0. & 1.000 \\
$\alpha_{Mm1}$ & Multiplicative tillage-cropping treatment effect for treatment Mm1. & 0.999 \\
$\alpha_{Ii0}$ & Multiplicative tillage-cropping treatment effect for treatment Ii0. & 0.999 \\
$\alpha_{Ii1}$ & Multiplicative tillage-cropping treatment effect for treatment Ii1. & 1.000 \\
$\alpha_{Im0}$ & Multiplicative tillage-cropping treatment effect for treatment Im0. & 1.000 \\
$\alpha_{Im1}$ & Multiplicative tillage-cropping treatment effect for treatment Im1. & 0.999 \\
$\alpha_{In0}$ & Multiplicative tillage-cropping treatment effect for treatment In0. & 1.000 \\
$\alpha_{In1}$ & Multiplicative tillage-cropping treatment effect for treatment In1. & 0.999 \\
$\alpha_{Mn0}$ & Multiplicative tillage-cropping treatment effect for treatment Mn0. & 1.000 \\
$\alpha_{Mn1}$ & Multiplicative tillage-cropping treatment effect for treatment Mm1. & 1.000 \\
$\alpha_{PF}$ & Multiplicative tillage-cropping treatment effect for treatment PF. & 1.000 \\
$\pi_{M \rightarrow D}$ & Proportion of manure to D pool & 1.000 \\
$\pi_{M \rightarrow R}$ & Proportion of manure to R pool & 1.000 \\
$\pi_{M \rightarrow F}$ & Proportion of manure to F pool & 1.000 \\
$\pi_{M \rightarrow S}$ & Proportion of manure to S pool & 0.999 \\
$\pi_{M \rightarrow H}$ & Proportion of manure to H pool & 1.000 \\
$p_{X \rightarrow F}$ & Proportion of soil-carbon from $X \in \{D, R, F, S \}$ to $F$. & 1.000 \\
$p_{H \rightarrow S}$ & Proportion of soil-carbon from $H$ to $S$. & 1.000 \\
$p_{clay}$ & Proportion of the soil that is clay &1.000  \\
$r_{DPM/RPM}$ & Ratio of decomposable to resistant carbon in plant material. & 0.999 \\
\hline
\end{tabular}
\end{center}
\end{table}

\begin{table}\caption{MCMC convergence diagnostics for process-error variance parameters in (\ref{transitionEqStoch}).}\label{convergenceTable2}
\begin{center}
\footnotesize
\begin{tabular}{ | p{1.5cm} | p{8.0cm} | p{2cm} | }
\hline
Parameter & Description & $\hat{R}$ \\
\hline 
\hline 
$\sigma^2_D$ &  Variance of additive process noise of $D$ pool.  & 1.000 \\
$\sigma^2_R$ &  Variance of additive process noise of $R$ pool.  & 1.000 \\
$\sigma^2_F$ &  Variance of additive process noise of $F$ pool.  & 1.000 \\
$\sigma^2_S$ &  Variance of additive process noise of $S$ pool.  & 1.000 \\
$\sigma^2_H$ &  Variance of additive process noise of $H$ pool.  & 1.000 \\
\hline
\end{tabular}
\end{center}
\end{table}

\begin{table}\caption{MCMC convergence diagnostics for error variances for measured carbon fractions.}\label{convergenceTable3}
\begin{center}
\footnotesize
\begin{tabular}{ | p{1.5cm} | p{8.0cm} | p{2cm} | }
\hline
Parameter & Description & $\hat{R}$ \\
\hline 
\hline 
$\sigma^2_{\textup{POC}}$ &  Measurement error variance for log(POC).  & 0.999 \\
$\sigma^2_{\textup{ROC}}$ &  Measurement error variance for  log(ROC). & 0.999 \\
$\sigma^2_{\textup{TOC}}$ &  Measurement error variance for log(TOC). & 1.000 \\
\hline
\end{tabular}
\end{center}
\end{table}

\FloatBarrier
\vfill \null \pagebreak

\setcounter{equation}{0}
\renewcommand{\theequation}{S7.\arabic{equation}}
\setcounter{figure}{0}
\renewcommand{\thefigure}{S7.\arabic{figure}}
\setcounter{table}{0}
\renewcommand{\thetable}{S7.\arabic{table}}
\setcounter{section}{0}
\renewcommand{\thesection}{S7.\arabic{section}}

\section*{\LARGE{Supplemental Material 7: Posterior Distributions of Parameters.}}

\begin{figure} 
\begin{center} 
\includegraphics[width=0.7\textwidth]{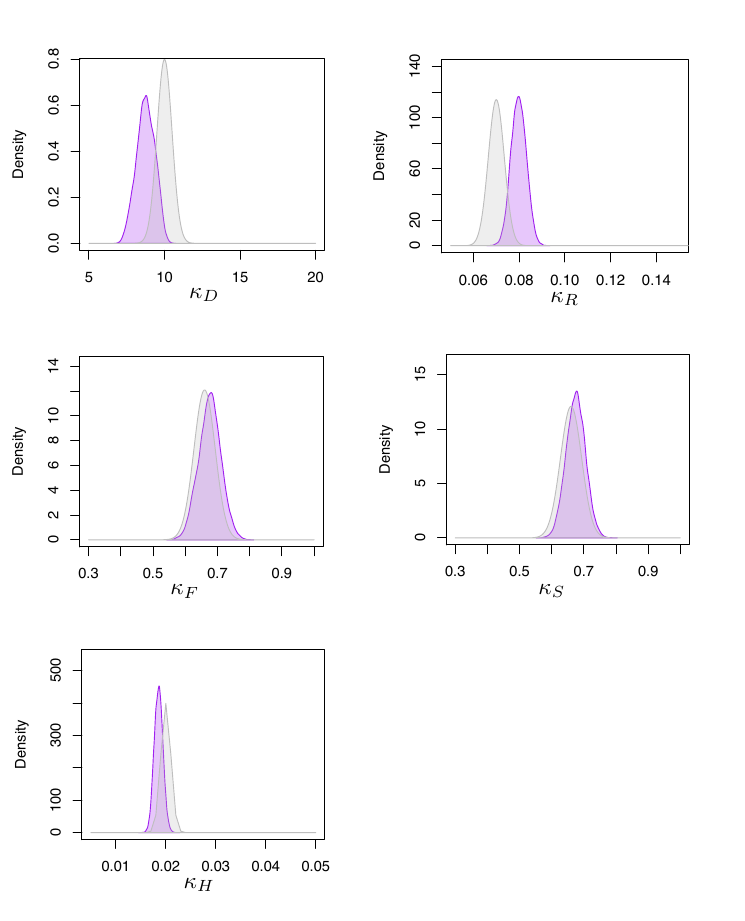}
\caption{Prior (gray) and posterior (purple) distributions for decay-rate parameters of CQUESST.  Plots give priority to the posterior distribution; when a prior is not visible, it indicates that it was very far from the posterior. \label{fig:priorpost1}}
\end{center} 
\end{figure}

\begin{figure} 
\begin{center} 
\includegraphics[width=0.7\textwidth]{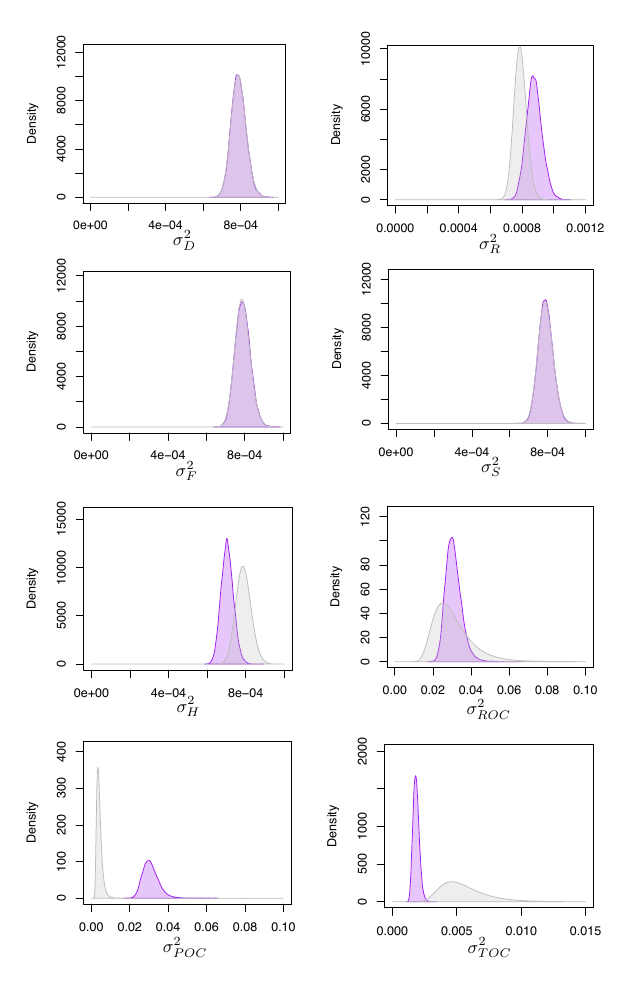}
\caption{Prior (gray) and posterior (purple) distributions for variance parameters of CQUESST. Plots give priority to the posterior distribution; when a prior is not visible, it indicates that it was very far from the posterior. \label{fig:priorpost2}}
\end{center} 
\end{figure}

\begin{figure} 
\begin{center} 
\includegraphics[width=0.7\textwidth]{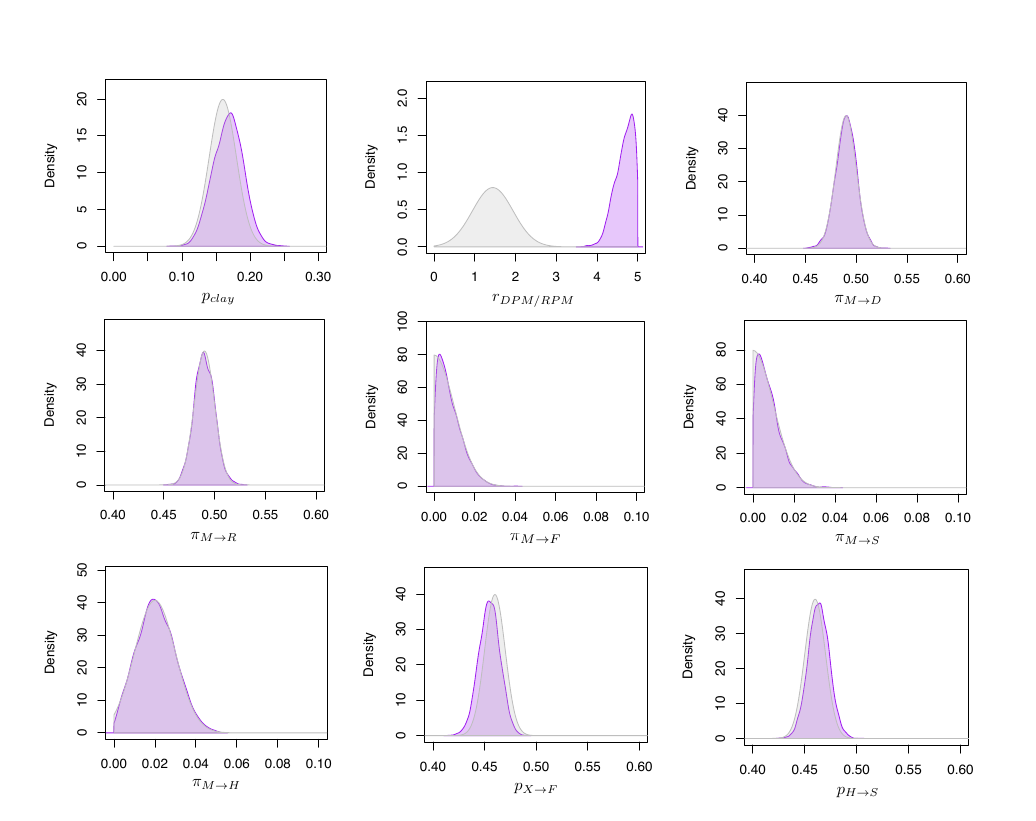}
\caption{Prior (gray) and posterior (purple) distributions for a subset of the parameters of CQUESST. Plots give priority to the posterior distribution; when a prior is not visible, it indicates that it was very far from the posterior. \label{fig:priorpost3}}
\end{center} 
\end{figure}

The CQUESST Bayesian hierarchical model was run in {\tt Stan}, from which we obtained posterior samples from six independent chains of the MCMC algorithm (see Supplemental Material 6 for details).  The samples from the three chains were aggregated into a larger set of samples and kernel density estimates were plotted alongside the prior distributions used in the parameter model.  These plots are provided in Figures \ref{fig:priorpost1}, \ref{fig:priorpost2}, and \ref{fig:priorpost3}.  In some cases, the posterior has deviated away from the prior (known as \emph{Bayesian learning}), indicating that the data has updated our beliefs about the distribution of the parameters.  In some other cases, the prior and posterior are very similar, indicating that the data has provided little additional information about the probability distribution of the parameter beyond what was contained in the prior.

\FloatBarrier
\vfill \null \pagebreak

\setcounter{equation}{0}
\renewcommand{\theequation}{S8.\arabic{equation}}
\setcounter{figure}{0}
\renewcommand{\thefigure}{S8.\arabic{figure}}
\setcounter{table}{0}
\renewcommand{\thetable}{S8.\arabic{table}}
\setcounter{section}{0}
\renewcommand{\thesection}{S8.\arabic{section}}

\section*{\LARGE{Supplemental Material 8: Analysis of Variance for the Millennium Tillage Trial.}}

\section{Examining MTT Treatment Main Effects.}
In order to compare CQUESST to a standard statistical approach, we undertook a classical analysis of variance (ANOVA) to examine how the treatments in the Millenium Tillage Trial affected total soil carbon (TOC).  This quantity was directly measured and hence an ANOVA is possible (CQUESST is able to make inferences on many other quantities that are out of the question for ANOVA).  

We used a linear mixed-effects model implemented using the {\tt glmmTMB} package for {\tt R} that included two fixed-effect factors: (i) the MTT treatments (having 14 factor levels); and (ii) observation months of the year (having 5 factor levels). A random effect was also included in the model for each of the 42 field-plots.  Consistent with the main text, we index factor levels for treatments, with $\tau \in \mathcal{T} = \{$PP, PF, Nn0, Nn1, Mm0, Mm1, Mn0, Mn1, In0, In1, Im0, Im1, Ii0, Ii1$\}$.  The factor levels for the month of observation were $\{$February, March, May, September, October$\}$.  

The linear model used in the ANOVA took the form:

\begin{equation}\label{linearModel}
Y_{i, j, k} = \mu + \beta_{\tau(i)} +  \gamma_j + \delta_i + \epsilon_{i,j,k}, \quad \textup{ where }  \delta_k \sim N(0, \sigma^2_{\delta}) \textup{ and }  \epsilon_l \sim N(0, \sigma^2_{\epsilon}).
\end{equation}

\noindent In (\ref{linearModel}) where: $Y_{i, j, k}$ is the $k$th measurement of TOC in the $i$th field-plot, undergoing treatment $\tau(i)$, where $\tau(i)$ is a function that returns the MTT treatment level for a given field-plot index $i$; $\mu$ is the grand mean; $\beta_{\tau(i)}$ is the effect of the $i$th level of the MTT treatment; $\gamma_j$ is the effect of the $j$th level of the observation month; $\delta_i$ is the random effect of field-plot $i$; and $\epsilon_{i,j,k}$ is the residual error.

\begin{figure} 
\begin{center}
\includegraphics[width=0.5\textwidth]{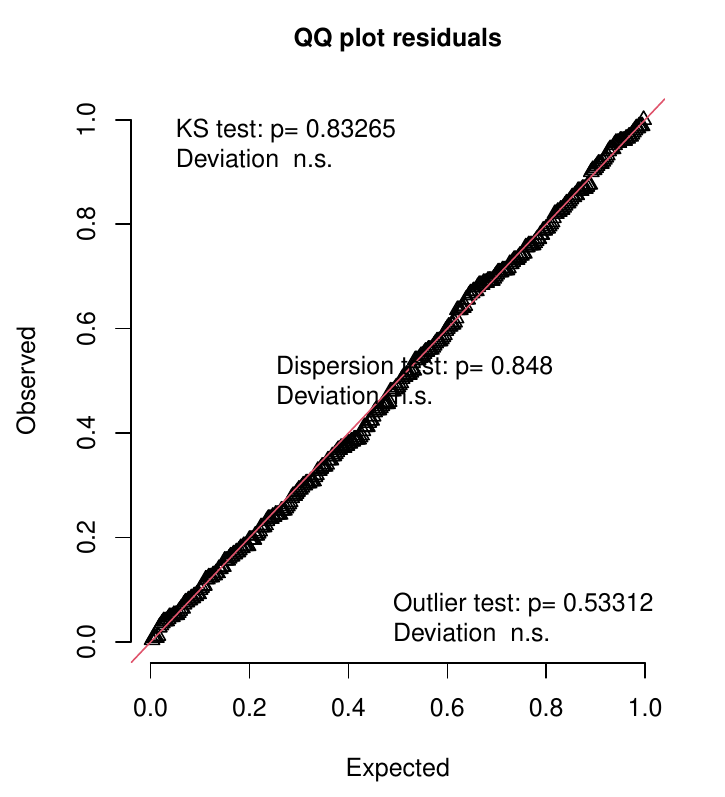}
\caption{Quantile-quantile diagnostic plot of residuals obtained for the model in Equation (\ref{linearModel}) using the {\tt DHARMa} package for {\tt R}, showing good agreement with the expected normal distribution.
\label{fig:qqplot}}
\end{center} 
\end{figure}

The fit of the model was assessed by diagnostics of the residuals using the {\tt DHARMa} package in {\tt R}.  Figure \ref{fig:qqplot} shows that the residuals closely follow the normal distribution (left-hand plot) and that the residuals were homoskedastic (right-hand plot).  Both of these attributes were assumptions made in formulating the model.

We used the {\tt emmeans} package in {\tt R} to construct contrasts that quantify the individual treatment effects ($\beta_{\tau(i)}$ in Equation (\ref{linearModel})) and test their statistical significance.  Table \ref{contrasts} provides these estimates and their associated $p$-values, with positive (negative) values of the treatment effect indicating that the treatment resulted in higher (lower) TOC on average.  Of note in Table \ref{contrasts} is that PP was the only treatment that showed a statistically significant deviation from the grand mean, which is consistent with  CQUESST inferences that PP was the only treatment showing strong evidence of sequestering carbon over the duration of the MTT (see Figure \ref{fig:carbon_change}).  

For comparison purposes, it is possible to compute a quantity similar to $\beta_{\tau(i)}$ for each MTT treatment using posterior estimates from CQUESST, of the latent TOC values.  We consider the quantity, 

\begin{equation}\label{beta_tilde}
\tilde{\beta}_{\tau} = \frac{1}{3} \sum_{j = 1}^3 ( \overline{TOC}_{i(\tau)_j} - \overline{TOC} ),
\end{equation}

\noindent where 

\begin{equation}
\overline{TOC}_{i(\tau)_j} \equiv \frac{1}{T} \sum_{t = 1}^T \sum_{X \in \{ D, R, F, S, H, I \}} X_{i(\tau)_j, t}
\end{equation}

\noindent is the posterior mean of total soil organic carbon over the duration of the MTT in the $j$th replicate of treatment $\tau \in \mathcal{T}$, $T$ is the total number of months over which the MTT was run;  and the quantity

\begin{equation}
\overline{TOC} = \frac{1}{42} \frac{1}{T} \sum_{i = 1}^{42} \sum_{t = 1}^T \sum_{X \in \{ D, R, F, S, H, I \}} X_{i, t}
\end{equation}

\noindent is the posterior mean of total soil organic carbon over the duration of the MTT across all field-plots.

\begin{figure} 
\begin{center}
\includegraphics[width=0.9\textwidth]{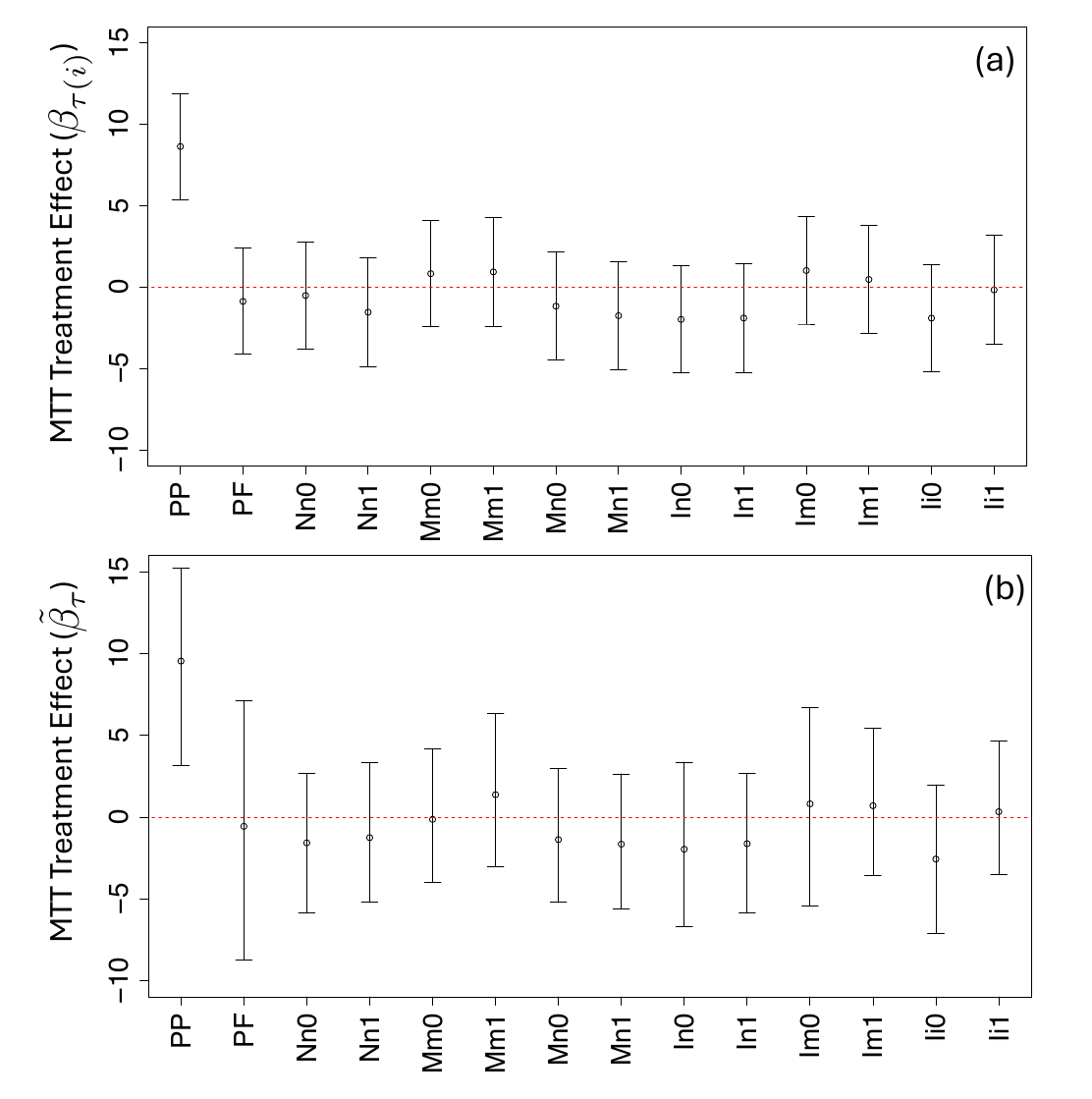}
\caption{Plot of estimated treatment effects for the MTT obtained from: (a) the individual treatment effects $\beta_{\tau(i)}$ in the linear ANOVA model (\ref{linearModel}); and (b) the individual treatment effects $\tilde{\beta}_{\tau}$ (given by (\ref{beta_tilde})) in the latent soil-carbon pools in CQUESST.}
\label{fig:comparison_plot}
\end{center} 
\end{figure}

CQUESST provides a posterior distribution of $\tilde{\beta}_{\tau}$, since this quantity can be computed for each sample made in the Markov Chain Monte Carlo (MCMC) algorithm.  Figure \ref{fig:comparison_plot} (a) shows the estimated treatment effects $\beta_{\tau(i)}$ and 95\% confidence intervals from the ANOVA.  Analogously, Figure \ref{fig:comparison_plot} (b) shows the posterior mean and 95\% credible intervals obtained for $\tilde{\beta}_{\tau}$ from CQUESST.  We note that both the ANOVA and CQUESST results show similar findings, namely, that the mean deviation from the grand mean was only different from zero for the permanent pasture (PP) treatment.

We remark that unlike the ANOVA, CQUESST can be used to make inferences on \emph{any} function of latent state variables and parameters in the model, not just mean deviations.  The inferences for all these quantities follow straight-forwardly from the MCMC samples.  As an example, Figure \ref{fig:carbon_change} in the main text provides posterior quantiles of carbon fluxes over the duration of the MTT, and the posterior probabilities that each treatment led to soil carbon sequestration.  In the same analysis, CQUESST also provides inferences such as in Figure \ref{fig:alphas} in the main text that show the relative changes in soil-carbon decay rates as a function of the MTT treatments.  As we discuss in the main text, this latter inference is particularly useful since it provides parameter estimates that can be used in mechanistic models to model heterogeneity in tillage and cropping practices, and ultimately to improve soil-carbon accounting.

Whilst an ANOVA is a much simpler model to fit, it can only be implemented on MTT observations, here TOC.  In contrast, CQUESST provides a more science-driven, mechanistic model to account for the variability in the MTT data and assimilates data from three types of observation (TOC, POC, and ROC) to make inferences on six latent soil-carbon pools per field-plot.  The relative sophistication of CQUESST for making inferences is highlighted when we compare Equations (\ref{transitionEqStoch}) and  (\ref{datamodel}) in the main text to the much simpler linear model in Equation (\ref{linearModel}).  Importantly, Equation (\ref{transitionEqStoch}) allows scientific knowledge about the way carbon cycles in the soil to be included in CQUESST, rather than being ignored.

\begin{table}\caption{Estimated treatment effects for the MTT.}\label{contrasts}
\begin{center}
\footnotesize
\begin{tabular}{ | c | c | c | c | c | c |}
\hline
Treatment Effect  & Estimate & Standard & Degrees of  & $t$-statistic & $p$-value \\
($\beta_i$) & & Error & Freedom & & (two-sided) \\
\hline 
Ii0 & -1.911 & 1.67 & 319 & -1.146 & 0.2526 \\
Ii1 & -0.174 & 1.70 & 319 & -0.102 & 0.9186 \\
Im0 & 1.016 & 1.68 &319  & 0.604 & 0.5460 \\
Im1 & 0.463 & 1.70 & 319 & 0.272 & 0.7856 \\
In0 & -1.991 & 1.68 & 319 & -1.184 & 0.2372 \\
In1 & -1.906 & 1.70 & 319 & -1.122 & 0.2629 \\
Mm0 & 0.820 & 1.67 & 319 & 0.491 & 0.6234 \\
Mm1 & 0.930 & 1.70 & 319 & 0.547 & 0.5845 \\
Mn0 & -1.175 & 1.69 & 319 & -0.697 & 0.4865 \\
Mn1 & -1.760 & 1.69 & 319 & -1.040 & 0.2992 \\
Nn0 & -0.520 & 1.67 & 319 & -0.312 & 0.7553 \\
Nn1 & -1.545 & 1.70 & 319 & -0.909 & 0.3638 \\
PF & -0.880 & 1.66 & 319 & -0.531 & 0.5955 \\
PP & 8.633 & 1.66 & 319 & \bf{5.213} & $\mathbf{<}$ \bf{0.0001} \\
\hline
\end{tabular}
\end{center}
\end{table}

\section{Examining Carbon Flux.}

Additionally, we constructed a linear model to examine changes in total organic carbon (TOC) over the duration of the Millennium Tillage Trial (MTT).  Equation (\ref{linearModel}) was extended to include a temporal trend:

\begin{equation}\label{linearModel_flux}
Y_{i, j, k} = \mu + \lambda_{\tau(i)}t_{j,k} + \beta_{\tau(i)} +  \gamma_j + \delta_i + \epsilon_{i,j,k}, \quad \textup{ where }  \delta_k \sim N(0, \sigma^2_{\delta}) \textup{ and }  \epsilon_l \sim N(0, \sigma^2_{\epsilon}).
\end{equation}

\begin{figure} 
\begin{center}
\includegraphics[width=0.5\textwidth]{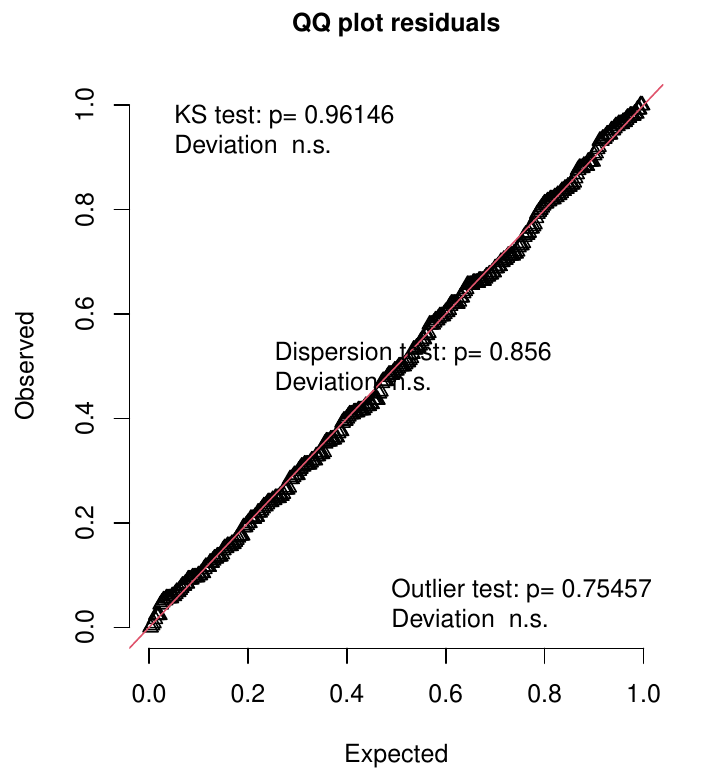}
\caption{Quantile-quantile diagnostic plot of residuals obtained for the model in Equation (\ref{linearModel_flux}) using the {\tt DHARMa} package for {\tt R}, featuring three tests showing good agreement with a modeled normal distribution.
\label{fig:qqplot2}}
\end{center} 
\end{figure}

\noindent In (\ref{linearModel_flux}) $\lambda_{\tau(i)}$ is the rate of change in TOC (measured in tonnes per hectare per month) for treatment ${\tau(i)}$ in field-plot $i$, and $t_{j,k}$ denotes the number of months elapsed in the MTT at the time of the $k$th observation in field-plot $j$.  Figure \ref{fig:qqplot2} provides a diagnostic assessment of the model's residuals, showing that they follow a normal distribution and are homoskedastic.  Table \ref{lambda_estimates} shows the estimated rates of change and significance values from the fitted model.  We see that permanent fallow (PF) shows strong evidence of a negative rate of change, indicating that TOC has decreased over time.  There is also strong evidence that TOC under treatment Nn0 declined over the duration of the MTT.  There is also some weaker evidence that treatment Mm1 shows evidence of TOC decreasing over time, whilst TOC under PP increased over time.

\begin{table}\caption{Estimates of rate of change $\lambda_{\tau(i)}$ in the fitted linear model given by Equation (\ref{linearModel_flux}).}\label{lambda_estimates}
\begin{center}
\footnotesize
\begin{tabular}{ | c | c | c | c | c |}
\hline
Treatment Effect  & Estimate & Standard & $z$-statistic & $p$-value \\
 & & Error & & (two-sided) \\
\hline 
$\lambda_{\textup{Ii0}}$ & 7.827e-03  & 2.281e-02 & 0.34 & 0.731530 \\
$\lambda_{\textup{Ii1}}$ & -2.217e-02  & 2.500e-02 & -0.89 & 0.375214 \\
$\lambda_{\textup{Im0}}$ & 3.064e-02 & 2.374e-02 &  1.29 & 0.196864 \\
$\lambda_{\textup{Im1}}$ & 1.593e-06 & 2.500e-02 &  0.00 & 0.999949 \\
$\lambda_{\textup{In0}}$ & 2.482e-02 & 2.374e-02 & 1.05 & 0.295804 \\
$\lambda_{\textup{In1}}$ & 5.448e-03 & 2.500e-02 & 0.22 & 0.827528 \\
$\lambda_{\textup{Mm0}}$ & -3.045e-02 & 2.281e-02 & -1.33 & 0.181880 \\
$\lambda_{\textup{Mm1}}$ & -4.246e-02 & 2.500e-02 & -1.70 & 0.089484 \\
$\lambda_{\textup{Mn0}}$ & -7.287e-03 & 2.405e-02 & -0.30 & 0.761887 \\
$\lambda_{\textup{Mn1}}$ & -2.593e-02 & 2.446e-02 & -1.06 & 0.289149 \\
$\lambda_{\textup{Nn0}}$ & -5.791e-02 & 2.281e-02 & \bf{-2.54} & \bf{0.011138} \\
$\lambda_{\textup{Nn1}}$ & -3.591e-02 & 2.500e-02 & -1.44 & 0.151005 \\
$\lambda_{\textup{PF}}$ & -1.222e-01 & 2.165e-02 & \bf{-5.65} & \bf{1.64e-08} \\
$\lambda_{\textup{PP}}$ & 3.249e-02 & 2.165e-02 & 1.50 & 0.133409 \\
\hline
\end{tabular}
\end{center}
\end{table}

\begin{figure} 
\begin{center}
\includegraphics[width=0.9\textwidth]{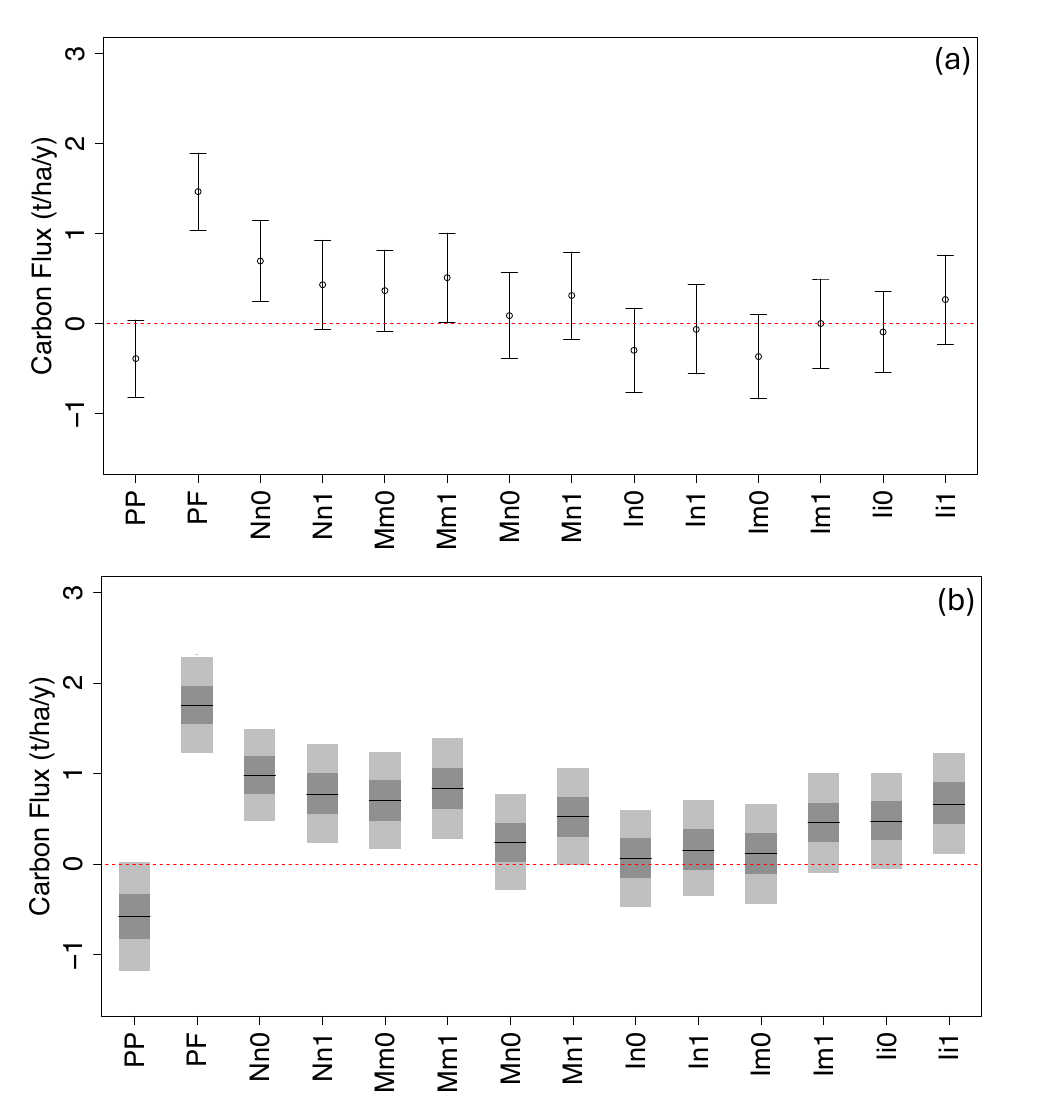}
\caption{Estimated carbon fluxes for MTT treatments using: (a) the linear model given by  Equation (\ref{linearModel_flux}) with circles showing the estimated mean and vertical bars spanning the 90\% confidence interval; and (b) CQUESST, with the horizontal line showing the median of the posterior distribution, dark ribbons spanning the 50\% credible interval and lighter bars spanning the 90\% credible interval.}
\label{fig:flux_anova}
\end{center} 
\end{figure}

Using the values of $\lambda_{\tau(i)}$, we can estimate the expected carbon flux from the soil in tonnes of carbon per hectare per year as $A_{\tau} = -12 \lambda_{\tau}$.  Figure \ref{fig:flux_anova} shows the estimated flux from the linear model besides the fluxes that were estimated using CQUESST.  There are obvious similarities between Figures \ref{fig:flux_anova}(a) and (b) suggesting that both the linear model with trend and CQUEST provided similar results for sequestered TOC.  It is noteworthy that CQUESST estimated noticeably larger fluxes for most of the treatments other than PP and that several of the 90\% credible intervals from CQUESST did not contain zero whilst the analogous 90\% confidence intervals did.  This comparison suggests that whilst CQUESST and the linear model can both be used to estimate carbon fluxes, CQUESST may provide greater statistical power for detecting carbon sequestration differences where they exist.  This is not surprising since CQUESST assimilates data from three observation types (TOC, POC, and ROC), and it is based on a scientifically motivated dynamical, nonlinear process model of soil-carbon cycling.  In contrast, the model presented in Equation (\ref{linearModel_flux}) is linear, does not use descriptives and only draws upon the observations of TOC.

\FloatBarrier
\vfill \null \pagebreak

\bibliographystyle{imsart-nameyear} 
\bibliography{References}

\end{document}